\documentclass[aps, twocolumn, showpacs, amsmath,superscriptaddress]{revtex4}
\usepackage{dcolumn}
\usepackage{bm}
\usepackage{graphicx}
\usepackage{color}
\usepackage{ulem}
\DeclareGraphicsExtensions{. jpg,. pdf, . mps, . png, . eps, . ps, . EPS}

\begin{document}
\def\be{\begin{equation}}
\def\ee{\end{equation}}

\def\bc{\begin{center}} 
\def\ec{\end{center}}
\def\bea{\begin{eqnarray}}
\def\eea{\end{eqnarray}}
\newcommand{\avg}[1]{\langle{#1}\rangle}
\newcommand{\ket}[1]{\left |{#1}\right \rangle}
\newcommand{\beq}{\begin{equation}}
\newcommand{\eneq}{\end{equation}}
\newcommand{\beqnn}{\begin{equation*}}
\newcommand{\eneqnn}{\end{equation*}}
\newcommand{\beqy}{\begin{eqnarray}}
\newcommand{\eneqy}{\end{eqnarray}}
\newcommand{\beqynn}{\begin{eqnarray*}}
\newcommand{\eneqynn}{\end{eqnarray*}}
\newcommand{\half}{\mbox{$\textstyle \frac{1}{2}$}}
\newcommand{\proj}[1]{\ket{#1}\bra{#1}}
\newcommand{\av}[1]{\langle #1\rangle}
\newcommand{\braket}[2]{\langle #1 | #2\rangle}
\newcommand{\bra}[1]{\langle #1 | }
\newcommand{\Avg}[1]{\left\langle{#1}\right\rangle}
\newcommand{\inprod}[2]{\braket{#1}{#2}}
\newcommand{\upket}{\ket{\uparrow}}
\newcommand{\downket}{\ket{\downarrow}}
\newcommand{\Tr}{\mathrm{Tr}}
\newcommand{\hcontrol}{*!<0em,.025em>-=-{\Diamond}}
\newcommand{\hctrl}[1]{\hcontrol \qwx[#1] \qw}
\newenvironment{proof}[1][Proof]{\noindent\textbf{#1.} }{\ \rule{0.5em}{0.5em}}
\newtheorem{mytheorem}{Theorem}
\newtheorem{mylemma}{Lemma}
\newtheorem{mycorollary}{Corollary}
\newtheorem{myproposition}{Proposition}
\newcommand{\vp}{\vec{p}}
\newcommand{\Or}{\mathcal{O}}
\newcommand{\so}[1]{{\ignore{#1}}}

\newcommand{\red}[1]{\textcolor{red}{#1}}
\newcommand{\blue}[1]{\textcolor{blue}{#1}}

\title{Statistical Mechanics of Multiplex Ensembles: Entropy and Overlap }

\author{Ginestra Bianconi}

\affiliation{School of Mathematical Sciences, Queen Mary University of London, London E1 4NS, United Kingdom}

\begin{abstract}

There is growing interest in multiplex networks where individual nodes take part in several layers of networks  simultaneously. This is the case for example in social networks where each individual node has different kind of social ties or  transportation systems where each location is connected to another location by different types of transport. Many of these multiplex are characterized by a significant overlap of the links in different layers. In this paper we introduce a statistical mechanics framework to describe multiplex ensembles. A multiplex is a system formed by $N$ nodes and $M$ layers of interactions where each node belongs to the $M$ layers at the same time. Each layer $\alpha$ is formed by a network $G^{\alpha}$. Here we introduce the concept of correlated multiplex ensembles in which the existence of a link in one layer is correlated with the existence of a link in another layer. This implies that a typical multiplex of the ensemble can have   a significant overlap of the links in the different layers. Moreover we characterize microcanonical and canonical multiplex ensembles satisfying respectively hard and soft constraints and we discuss how to construct multiplex in these ensembles. Finally we  provide the  expression for the entropy of these ensembles that can be useful to address different inference problems involving multiplexes. 
\end{abstract}


\maketitle
\section{Introduction}

In the last years large attention has been paid  to single networks 
 \cite{RMP,Newman_rev,Boccaletti2006,Fortunato} with  breakthroughs revealing  the deep relation between topological properties of the networks and their dynamics \cite{crit,Dynamics}. 
Nevertheless, many systems  are not formed by isolated networks, instead they are formed by a network
of networks \cite{Mucha,Thurner,Havlin3}.  Examples include multimodal transportation networks~\cite{Kurant2006, traffic}, climatic
systems~\cite{Donges2011}, economic markets~\cite{Yang2009},
energy-supply networks~\cite{Havlin1} and the human
brain~\cite{Bullmore2009}.  Moreover many networks are multiplex indicating
the fact that two nodes can belong to different networks at the same time.
For example this is the case of social networks in which agents can be linked at  same time, by familiar relationships,
friendship, professional collaboration, co-location, email communication and so on. The offshoot of the network theory fundamental insights  is that for us
working in statistical mechanics it is now possible - in a sense it is mandatory – to move into the field to shed light on the complexity on interdependent networks and multiplexes. 
In this context, new measures for multiplex \cite{Thurner,Boccaletti,Marc} and new models of  growing multiplexes \cite{Growth1, Growth2} have been proposed. Moreover, several works have studied dynamical processes taking place on multiplexes and interacting networks and new surprising phenomena have been observed in this context involving percolation \cite{Havlin1, Havlin2,Son,JSTAT}, cascades  \cite{Leicht}, diffusion  \cite{Diffusion}, epidemic spreading \cite{Boguna} and cooperation  \cite{Cooperation},opinion dynamics \cite{EPL} and community detection\cite{Mucha,N1,N2}.

Yet, we are only at the beginning of the research on interacting networks and multiplexes and we  need to develop further theoretical frameworks to extract information from multiplex data. For this purpose we need new statistical mechanics methods to analyze multiplex and interacting networks data.

An important tool to study real networks is to compare them with null models represented by randomized network ensembles.
For single networks an equilibrium statistical mechanics framework has been recently formulated  \cite{Newman1,Newman2,BianconiEPL,BC,AB2009,AB2010,Annibale,Ne1,NeK2,Munoz,Garlaschelli,Peixoto,Bornholdt} in order to characterize network ensembles. 
A network ensemble is defined as a set of networks that satisfy a given number of structural
constraints, i.e. degree sequence, community structure etc. Every set of constraints can
give rise to a {\it microcanonical network ensemble}, satisfying the hard constraints, or to a
{\it canonical network ensemble} in which the constraints are satisfied in average. This
construction is symmetric to the classical ensemble in statistical mechanics where one
considers system configurations compatible either with a fixed value of the energy
(microcanonical ensembles) or with a fixed average of the energy determined by the thermal
bath (canonical ensembles). For example the $G(N,L)$ random graphs formed by networks of
$N$ nodes and $L$ links is an example of microcanonical network ensemble while the $G(N,p)$
ensembles, where each pair of links is connected with probability $p$, is an example of
canonical network ensemble since the number of links can fluctuate but has a fixed average
given by $\avg{L}=pN(N-1)/2$. A theoretical question that arise in the study of network ensembles is whether the microcanonical ensemble and the corresponding canonical ensemble are equivalent in the thermodynamics limit.  It turns out \cite{AB2009,AB2010} that when the number of constraints in two conjugated
network ensembles is extensive, the ensembles are no longer equivalent in the
thermodynamic limit and it is important to characterize their differences. For example microcanonical and canonical network ensembles with
given degree sequence are non equivalent in the thermodynamic limit. 

The entropy of network ensembles is given by the logarithm of the number of typical networks in the ensemble. The entropy of a network ensemble  quantifies the complexity of the ensemble. In particular we have that the smaller is
the entropy of the ensemble the smaller is the number of networks satisfying the
corresponding constraints and implying that these networks  are more optimized.
Both the network ensembles and their entropy can be used on several inference problems to
extract information from a given network  \cite{Leicht2,PNAS} . Given the relevance of the statistical mechanics of randomized network ensembles for describing real networks, it is important to extend this successful approach to describe multiplex ensembles.  In this paper we have chosen to consider only simple multiplex but the results can be easily extended to directed and weighted networks. We plan to consider these more complex cases in later publications.
 
In this paper we will show how to treat multiplex ensembles as null models for multiplexes. We will introduce a distinction between uncorrelated multiplex ensembles and correlated multiplex ensembles in which the existence of a link in one layer is correlated to the existence of a link in another layer. We will characterize the overlap between links in two different layers in the case of uncorrelated and correlated multiplex ensembles. We will evaluate the entropy of microcanonical and canonical multiplex ensembles for a large variety of constraints. Finally this work  open a new scenario for building null models of multiplex ensembles that has the promise to be used in a large variety of inference problems.
The paper is organized as follows.
In section II we introduce multiplexes and the global and local overlap of the links in the two layers.
In section III we introduce multiplex ensembles, their entropy and correlations.
In section IV  we describe canonical multiplex ensembles, we distinguish these ensembles as correlated or uncorrelated. We give relevant examples of these ensembles, we calculate their entropy and their overlap and give algorithms to construct multiplexes in these ensembles.
In section V we describe microcanonical multiplex ensembles. We give relevant examples of both correlated and uncorrelated microcanonical multiplex ensembles and calculate their entropy.
Finally in section VI we make the concluding remarks.
  
\section{Multiplex and overlap between two layers}
\label{ml}

Consider a multiplex  formed by $N$ labelled nodes $i=1,2\ldots, N$
and $M$ layers. We can represent the multiplex as described for example in \cite{Mucha}. To this end we indicate by $\vec{G}=(G^1,G^2\ldots G^M)$ the set of all the networks $G^{\alpha}$ at layer $\alpha=1,2,\ldots, M $ forming the multiplex. Each of  these networks has an  adjacency matrix
with matrix elements $a_{ij}^{\alpha}=1$ if there is a link between node $i$ and node $j$ in layer $\alpha=1,2,\ldots M $ and zero otherwise. 
Moreover for a multiplex we can define {\it multilinks}, and {\it multidegrees} in the following way.
Let us consider the  vector $\vec{m}=(m_1,m_2,\ldots,m_{\alpha},\ldots m_M)$ in which every element $m_{\alpha}$ can take only two values $m_{\alpha}=0,1$.
We define a {\it multilink} $\vec{m}$ the set of links connecting a given pair of nodes in the different layers  of the multiplex and connecting them in the generic  layer $\alpha$ only if $m_{\alpha}=1$.
We can therefore introduce the {\it multiadjacency matrices $A^{\vec{m}}$}  with elements
$A_{ij}^{\vec{m}}$ equal to 1 if there is a multilink $\vec{m}$ between node $i$ and node $j$ and zero otherwise, i.e.
the multiadjacency matrices have elements $A_{ij}^{\vec{m}}=0,1$ given by 
  \begin{equation}
 A_{ij}^{\vec{m}}=\prod_{\alpha=1}^M \left[a_{ij}^{\alpha}m_{\alpha}+(1-a_{ij}^{\alpha})(1-m_{\alpha})\right].
 \label{MA}
 \end{equation}
Therefore we can define the total number of multilinks $\vec{m}$ in a network as the total number of pairs of nodes connected by a multilink $\vec{m}$. Moreover we can define the {\it multidegree} $\vec{m}$  of a node $i$, $k_i^{\vec{m}}$ as the total number of multilinks $\vec{m}$ incident to node $i$, i.e.
\begin{equation}
k_i^{\vec{m}}=\sum_{j=1}^N A_{ij}^{\vec{m}}.
\end{equation}
We note here that the multilink $\vec{m}=\vec{0}$ between two nodes represent the situation in which in all the layers of the multiplex the two nodes are not directly linked. To have a uniform notation we refer also in this case to a multilink. Moreover we observe that the multiadjacency matrices  are not all independent. In fact they satisfy the following normalization condition
\begin{eqnarray}
\sum_{\vec{m}} A_{ij}^{\vec{m}}=1,
\end{eqnarray}
for every fixed pair of nodes $(i,j)$.

For two layers $\alpha,\alpha'$ of the multiplex  we can define the {\it global overlap} $O^{\alpha,\alpha'}$ as the total number of pair of nodes connected at the same time by a link in layer $\alpha$ and a link in layer $\alpha'$, i.e. 
\begin{equation}
O^{\alpha,\alpha'}=\sum_{i<j} a_{ij}^{\alpha}a_{ij}^{\alpha'}.
\label{Og}
\end{equation}
For a node $i$ of the multiplex, we can define the {\it local overlap} $o_i^{\alpha,\alpha'}$ of the links in two layers $\alpha$ and $\alpha'$ as the total number of nodes $j$ linked to the node $i$ at the same time by a link in layer $\alpha$ and a link in layer $\alpha'$, i.e.
\begin{equation}
o_i^{\alpha,\alpha'}=\sum_{j=1}^N a_{ij}^{\alpha}a_{ij}^{\alpha'}.
\label{oi}
\end{equation}
We expect the  global or the local overlap between two layers to characterize important correlations between the two layers in real-world situations.
For example in a transportation multiplex, where the different layers can represent different kind  of transport such as bus and  train connections or private commuting, we expect that the links in the different layers of this multiplex have an overlap which is statistically significant respect to a null hypothesis of uncorrelation between the different layers. Also in social sciences if we consider the multiplex formed by different means of communication between people, (emails, mobile, sms, etc.) two people that are linked in one layer are also likely to be linked in another layer, forming a multiplex of correlated networks.
We note also that for two layer multiplex, i.e.  $M=2$ the multilink $k^{1,1}_i$ is equal to the local overlap $o_i$. Reversibly, the multidegree $k_i^{\vec{m}}$ of a node $i$ in a multiplex with generic number of layers $M$ can be seen as a higher order local overlap.

\section{Multiplex ensembles, entropy and correlations}
A multiplex ensemble is specified when the probability $P(\vec{G})$ for each possible multiplex is given.
In a multiplex ensemble, if the probability of a multiplex is given by $P(\vec{G})$, the entropy of the multiplex $S$  is defined as 
\begin{equation}
S=-\sum_{\vec{G}}P(\vec{G})\log P(\vec{G}).
\label{entropy}
\end{equation}
and measures the logarithm of the typical number of multiplexes in the ensemble.
As it occurs for single networks we can construct microcanonical or canonical multiplex ensembles according to the equilibrium statistical mechanics approach applied to complex networks.
Moreover two layers in a multiplex network ensemble might be either correlated or uncorrelated.
We will say that a multiplex ensemble is {\it  uncorrelated} if  the probability $P(\vec{G})$ of the multiplex is factorizable into the probability of each single network $G^{\alpha}$ in the layer $\alpha$.
Therefore in an uncorrelated multiplex ensemble we have 
\begin{equation}
P(\vec{G})=\prod_{\alpha=1}^M P_{\alpha}(G^{\alpha}).
\label{uc0}
\end{equation}
where $P_{\alpha} (G^{\alpha})$ is the probability of network $G^{\alpha}$
on layer $\alpha$.
If Eq. (\ref{uc0})  doesn't hold,i.e.
\begin{equation}
P(\vec{G})\neq \prod_{\alpha=1}^M P_{\alpha}(G^{\alpha}).
\end{equation}
we will say that the multiplex ensemble is {\it correlated}.

Using Eq. $(\ref{uc0})$ we can show that the  entropy of any  uncorrelated multiplex ensemble is given by 
\begin{equation}
S=\sum_{\alpha=1}^M S^{\alpha}=-\sum_{\alpha=1}^M P_{\alpha}(G^{\alpha})\log P_{\alpha}(G^{\alpha}),
\end{equation}
where $S^{\alpha}$ is the entropy of the network ensemble in layer $\alpha$ with probability $P_{\alpha}(G^{\alpha})$.
 In an uncorrelated multiplex the links in any two layer  $\alpha$ and $\alpha'$  are uncorrelated  therefore we have 
\begin{equation}
\avg{a_{ij}^{\alpha}a_{ij}^{\alpha'}}=\avg{a_{ij}^{\alpha}}\avg{a_{ij}^{\alpha'}}
\end{equation}
for every choice of  pair of nodes $i,j$.

On the contrary if the multiplex is correlated there will be at least two layers $\alpha$ and $\alpha'$ in a multiplex ensemble and a pair of nodes $i$ and $j$ for which
\begin{equation}
\avg{a_{ij}^{\alpha}a_{ij}^{\alpha'}}\neq\avg{a_{ij}^{\alpha}}\avg{a_{ij}^{\alpha'}}.
\end{equation}

\section{Canonical multiplex ensembles or exponential random multiplexes}

The canonical multiplex ensembles are the set of multiplex that satisfy a series of constraints in average.

The construction of the canonical multiplex ensembles or exponential random multiplex follow closely the derivation or the exponential random graphs.

We can build a canonical multiplex ensemble by maximizing the entropy of the ensemble given by Eq. ($\ref{entropy}$) under the condition that the soft constraints we want to impose are satisfied.
We assume to have $K$ of such constraints determined by the conditions
\begin{equation}
\sum_{\vec{G}}P(\vec{G})F_{\mu}(\vec{G})=C_{\mu}
\label{constraints}
\end{equation}
for $\mu=1,2\ldots, K$, where $F_{\mu}(\vec{G})$ determines one of the  structural constraints that we want to impose to the network. For example, $F_{\mu}(\vec{G})$ might characterize  the total number of links in a layer of  the multiplex $\vec{G}$ or the degree of a node in a layer of the multiplex $\vec{G}$ etc.. In the following we will specify in detail different major examples for the constraints $F_{\mu}(\vec{G})$.
In order the build the maximal entropy ensemble satisfying the soft constraints defined Eqs. $(\ref{constraints})$, we  maximize the entropy $S$ given by Eq. $(\ref{entropy})$ under the condition that the ensemble satisfies  the $K$ soft constraints given by Eqs. $(\ref{constraints})$.
Introducing the Lagrangian multipliers $\lambda_{\mu}$ enforcing the conditions given by Eqs. $(\ref{constraints})$  and the Lagrangian multiplier $\Lambda$ enforcing the normalization of the probabilities $\sum_{\vec{G}}P(\vec{G})=1$ we find the expression for the probability $P(\vec{G})$ of a multiplex by solving the 
following system of equations, 
\begin{equation}
\frac{\partial }{\partial P(\vec{G})}\left[S-\sum_{\mu=1}^P \lambda_{\mu} \sum_{\vec{G}}F_{\mu}(\vec{G})P(\vec{G})-\Lambda\sum_{\vec{G}}P(\vec{G})\right]=0.
\end{equation}
Therefore we get  that the probability of a multiplex $P_C(\vec{G})$ in a canonical multiplex ensemble is given by 
\begin{equation}
P_C(\vec{G})=\frac{1}{Z_C}\exp\left[-\sum_{\mu}\lambda_{\mu}F_{\mu}(\vec{G})\right]
\label{PC}
\end{equation}
where the normalization constant $Z_C$ is called the  ``partition function " of the canonical multiplex ensemble.
The values of the Lagrangian multipliers $\lambda_{\mu}$ are  determined  by imposing the constraints given by Eq. $(\ref{constraints})$ assuming for the probability $P_C(\vec{G})$ the structural form given by Eq. $(\ref{PC})$.

In this ensemble, we can the relate the entropy $S$ (given by Eq. $(\ref{entropy})$)  to the canonical partition function $Z_C$ getting
\begin{eqnarray}
S&=&-\sum_{\vec{G}} P_C(\vec{G}) \log  P_C(\vec{G})\nonumber \\&=&-\sum_{\vec{G}} P_C(\vec{G})[-\sum_{\mu} \lambda_{\mu} F_{\mu}(\vec{G})-\log(Z_C)]\nonumber \\
&=&\sum_{\mu} \lambda_{\mu} C_{\mu}+\log{Z_C}.
\label{Sc}
\end{eqnarray}
We call the entropy $S$ of the canonical multiplex ensemble the {\it Shannon entropy} of the ensemble.

\subsection{Uncorrelated or correlated canonical multiplex ensembles}
For a canonical uncorrelated multiplex ensemble in which each multiplex $\vec{G}$ has probability $P(\vec{G})$, we have that  Eq. $(\ref{uc0})$ is satisfied, i.e.
\begin{equation}
P_C(\vec{G})=\prod_{\alpha=1}^M P^{\alpha}_C(G^{\alpha}).
\label{uc1}
\end{equation}
where $P^{\alpha}_C (G^{\alpha})$ is the probability of network $G^{\alpha}$
on layer $\alpha$.
Given the structure of the probability $P_C(\vec{G})$ in the canonical multiplex ensemble given by Eq. ($\ref{PC}$), in order to have an uncorrelated multiplex the functions $F_{\mu}(\vec{G})$ should be  equal to  a linear combination of constraints $f_{\mu,\alpha}(G^{\alpha})$ on the networks $G^{\alpha}$ on a single layer $\alpha$, i.e.
\begin{equation}
F_{\mu}(\vec{G})=\sum_{\alpha=1}^M f_{\mu,\alpha}(G^{\alpha}).
\end{equation} 
A special case of this type of constraints is when each constraint depends  on a single network $G^{\alpha}$ in a  layer $\alpha$.
In this case  typical sets of constraints can be: the average total number of link in each layer, the expected degree sequence in each layer, the expected degree sequence and the expected community structure in each layer etc. Instead, in the case  in which the multiplex is  correlated, also quantities such as the expected overlap can be fixed. For a multiplex formed by two layers, we can therefore construct multiplex ensembles with expected total number of links in each layer and expected global overlap between the two layers, or with expected degree sequence and expected local overlap between the two layers etc.

We can therefore construct a large class of canonical uncorrelated and correlated multiplex ensembles enforcing a different number of constraints. Starting with a minimal number of constraints,  when we introduce further constraints in our ensemble we expect that the typical number of multiplexes that  satisfy the constraints will decrease, and therefore we expect that the entropy of the multiplex ensemble will decrease. Multiplex in network ensembles with smaller typical number of realizations are more complex and more optimized. Therefore the entropy of the multiplex can be used in solving inference problems and is a first principle measure  to quantify the complexity of the ensemble.
In the following we give some example of uncorrelated and correlated canonical multiplex ensembles.
 
 \subsection{Examples of uncorrelated canonical multiplex ensembles}
\label{eu}
\subsubsection{Multiplex ensemble with given expected  total number of links in each layer}
\label{muc1}
 We can fix the average number of links in each layer $\alpha$ to be equal to  $L^{\alpha}$. In this case we have $K=M$ constraints  in the system indicated with a label $\alpha=1,2,\ldots, M$. These constraints are given by 
 \begin{equation}
\sum_{\vec{G}} F_{\alpha}(\vec{G})P(\vec{G})=\sum_{\vec{G}}\sum_{i<j}a_{ij}^{\alpha}P(\vec{G})=L^{\alpha}
\label{1uc}
 \end{equation}
 with $\alpha=1,2,\ldots, M$.
Therefore the explicit expression for $F_{\alpha}(\vec{G})$ is given by
 \begin{equation}
 F_{\alpha}(\vec{G})=\sum_{i<j}a_{ij}^{\alpha}.
 \end{equation}
 The probability of the multiplex is given by Eq. $(\ref{PC})$. Using this expression we observe that the probability $P_C(\vec{G})$ can be written as 
 \begin{equation}
 P_C(\vec{G})=\frac{1}{Z_C}\exp\left[-\sum_{\alpha=1}^M\lambda_{\alpha}\sum_{i<j}a_{ij}^{\alpha}\right]
 \label{PCLU}
 \end{equation}
 where $Z_C$ is the canonical partition function and $\lambda_{\alpha}$ is the Lagrangian multiplier enforcing the constraint  given by Eq. $(\ref{1uc})$.
 The probability of a link between node $i$ and node $j$ in layer $\alpha$ is given by 
 \begin{equation}
 p_{ij}^{\alpha}=p^{\alpha}=\avg{a_{ij}^{\alpha}}=\frac{e^{-\lambda_{\alpha}}}{1+e^{-\lambda_{\alpha}}}
 \end{equation}
 The Lagrangian multipliers are fixed by the condition
 \begin{equation}
 \sum_{i<j}p_{ij}^{\alpha}=\frac{N(N-1)}{2}p^{\alpha}=L^{\alpha},
 \end{equation}
 i.e. $p^{\alpha}=2L^{\alpha}/[N(N-1)]$ and $e^{-\lambda_{\alpha}}=\frac{2L^{\alpha}}{{N(N-1)}-{2L^{\alpha}}}$.
 Using the definition of the entropy of the multiplex Eq. $(\ref{entropy})$ and the expression for $P_C(\vec{G})$ given by Eq. $(\ref{PCLU})$ it is easy to show that the entropy of the canonical multiplex ensemble $S$, that we call Shannon entropy, is given by  
 \begin{equation}
 S=-\frac{N(N-1)}{2}\sum_{\alpha=1}^M[p^{\alpha}\log p^{\alpha}+(1-p^{\alpha})\log(1-p^{\alpha})].
 \end{equation}
where  $p^{\alpha}=2L^{\alpha}/[N(N-1)]$. 
If the number of layers $M$ is finite, it can be shown that this expression in the large $N$ limit, is equal to 
\begin{equation}
S=\sum_{\alpha=1}^M\log \left(\begin{array}{c} \frac{N(N-1)}{2} \nonumber \\ L^{\alpha}\end{array}\right).
\end{equation}
 
 \subsubsection{Multiplex ensemble with given expected degree sequence in each layer}
 We can fix the expected degree $k_i^{\alpha}$ of every node $i$ in each layer $\alpha$.
 In this case we have $K=M\times N$ constraints  in the system indicated with a labels $\alpha=1,2,\ldots, M$ and $i=1,2\ldots, N$. These constraints are given by 
 \begin{equation}
\sum_{\vec{G}} F_{i,\alpha}(\vec{G})P(\vec{G})=\sum_{\vec{G}}\sum_{j=1,j\neq i}^Na_{ij}^{\alpha}P(\vec{G})=k_i^{\alpha}.
\label{2uc}
 \end{equation}
Therefore the explicit expression for $F_{i,\alpha}(\vec{G})$ is given by
 \begin{equation}
 F_{i,\alpha}(\vec{G})=\sum_{j=1,j\neq i}^Na_{ij}^{\alpha}.
 \end{equation}
 The probability of the multiplex is given by Eq. $(\ref{PC})$. Using this expression we observe that the probability $P_C(\vec{G})$ can be written as 
 \begin{equation}
 P_C(\vec{G})=\frac{1}{Z_C}\exp\left[-\sum_{\alpha=1}^M\sum_{i=1}^N\lambda_{i,\alpha}\sum_{j=1,j\neq i}^Na_{ij}^{\alpha}\right]
 \label{2PCUL}
 \end{equation}
 where $Z_C$ is the canonical partition function and $\lambda_{i,\alpha}$ is the Lagrangian multiplier enforcing the constraint  given by Eq. $(\ref{2uc})$.
 The probability of a link between node $i$ and node $j$ in layer $\alpha$ is given by 
 \begin{equation}
 p_{ij}^{\alpha}=\avg{a_{ij}^{\alpha}}=\frac{e^{-\lambda_{i,\alpha}-\lambda_{j,\alpha}}}{1+e^{-\lambda_{i,\alpha}-\lambda_{j,\alpha}}}
 \end{equation}
 where the Lagrangian multipliers $\lambda_{i,\alpha}$ are fixed by the conditions 
 \begin{equation}
 \sum_{j=1,j\neq i}^N p_{ij}^{\alpha}=k_i^{\alpha}.
 \end{equation}
 
 Using the definition of the entropy of the multiplex Eq. $(\ref{entropy})$ and the expression for $P_C(\vec{G})$ given by Eq. $(\ref{2PCUL})$ it is easy to show that the entropy of the canonical multiplex ensemble $S$, that we call Shannon entropy, is given by 
 \begin{equation}
 S=-\sum_{\alpha=1}^M\sum_{i<j}[p_{ij}^{\alpha}\log p_{ij}^{\alpha}+(1-p_{ij}^{\alpha})\log(1-p_{ij}^{\alpha})].
 \label{Suc2}
 \end{equation}
 If $k_i^{\alpha}<\sqrt{\avg{k^{\alpha}}N}\ \forall i=1,2\ldots, N$ then  each network $G^{\alpha}$ is uncorrelated and therefore,  $e^{-\lambda_{i,\alpha}}\simeq\frac{k_i^{\alpha}}{\sqrt{\avg{k^{\alpha}}N}}$ and $p_{ij}^{\alpha}\simeq \frac{k_i^{\alpha}k_j^{\alpha}}{\avg{k^{\alpha}}N}$. 
 In this limit the Shannon entropy $S$ is given by 
 \begin{eqnarray}
 S&\simeq&\sum_{\alpha=1}^P\left[-\sum_i k_i^{\alpha}\log(k_i^{\alpha})+\frac{1}{2}\avg{k^{\alpha}}N\log(\avg{k^{\alpha}}N)\right.\nonumber \\
&& \left.+\frac{1}{2}\avg{k^{\alpha}}N-\frac{1}{4}\left(\frac{\avg{(k^{\alpha})^2}}{\avg{k^{\alpha}}}\right)^2\right].
\label{Suc2b}
 \end{eqnarray}

 \subsubsection{Multiplex ensemble with given  expected number of links present in each layer between  nodes in different communities}
 We can fix the expected number of links  present in each layer between nodes belonging to different communities.
 We assign to each node $i$ a discrete variable $q_i=1,2,\ldots, Q$ indicating the community of the node. 
  We can consider canonical uncorrelated multiplex ensembles in which we  fix  the expected number of links $e_{q,q'}^{\alpha}$ between nodes in community $q$ and nodes in community $q'$ in layer $\alpha$.
 In this case we have $K=M\times Q(Q+1)/2$ constraints  in the system indicated with a labels $\alpha=1,2,\ldots, M$ and $q,q'=1,2\ldots, Q$. These constraints are given by 
 \begin{eqnarray}
\sum_{\vec{G}} F_{q,q'\alpha}(\vec{G})P(\vec{G})&=&e_{q,q'}^{\alpha}
\label{3uc}
 \end{eqnarray} 
where the explicit expression for $F_{q,q',\alpha}(\vec{G})$ is given by
 \begin{eqnarray}
 F_{q,q',\alpha}(G^{\alpha})&=&\sum_{i,j}a_{ij}^{\alpha}\delta_{q,q_i}\delta_{q',q_j}, \ \mbox{for} \  q\neq q' \nonumber \\
 F_{q,q,\alpha}(G^{\alpha})&=&\sum_{i<j}a_{ij}^{\alpha}\delta_{q,q_i}\delta_{q,q_j}.
 \end{eqnarray}
  The probability of the multiplex is given by Eq. $(\ref{PC})$. Using this expression we observe that the probability $P_C(\vec{G})$ can be written as 
 \begin{equation}
 P_C(\vec{G})=\frac{1}{Z_C}\exp\left[-\sum_{\alpha=1}^M\sum_{q\leq q'}\lambda_{q,q',\alpha} F_{q,q',\alpha}(\vec{G})\right]
 \label{3PCUL}
 \end{equation}
 where $Z_C$ is the canonical partition function and $\lambda_{q,q',\alpha}$ is the Lagrangian multiplier enforcing the constraint  given by Eq. ($\ref{3uc}$).
 The probability of a link between node $i$ and node $j$ in layer $\alpha$ is given by 
 \begin{equation}
 p_{ij}^{\alpha}=\avg{a_{ij}^{\alpha}}=\frac{e^{-\lambda_{q_i,q_j,\alpha}}}{1+e^{-\lambda_{q_i,q_j,\alpha}}}
 \label{pijbuc}
 \end{equation}
 where the Lagrangian multipliers are fixed by the conditions 
 \begin{eqnarray}
 \sum_{i,j} p_{ij}^{\alpha}\delta_{q,q_i}\delta_{q',q_j}&=&e_{q,q'}^{\alpha} \ \mbox{for}\ q\neq q'\nonumber \\
  \sum_{i<j} p_{ij}^{\alpha}\delta_{q,q_i}\delta_{q,q_j}&=&e_{q,q}^{\alpha}
 \end{eqnarray}
As it can be seen by Eq. $(\ref{pijbuc})$ the probabilities $p_{ij}^{\alpha}$ depend only on $q_i, q_j$ and $\alpha$ therefore we have $ p_{ij}^{\alpha}=p^{\alpha}(q_i,q_j)$ with
\begin{eqnarray}
p^{\alpha}(q,q')&=&\frac{e_{q,q'}^{\alpha}}{n_q n_{q'}} \ \mbox{for} \ q\neq q' \nonumber \\
p^{\alpha}(q,q)&=& \frac{e_{q,q}^{\alpha}}{n_q(n_q-1)/2}
\end{eqnarray}
where $n_q$ indicates the total number of nodes in community $q$.
 Using the definition of the entropy of the multiplex Eq. $(\ref{entropy})$ and the expression for $P_C(\vec{G})$ given by Eq. $(\ref{3PCUL})$ it is easy to show that the entropy of the canonical multiplex ensemble $S$, that we call Shannon entropy, is given by 
 \begin{equation}
 S=-\sum_{\alpha=1}^M\sum_{i<j}[p_{ij}^{\alpha}\log p_{ij}^{\alpha}+(1-p_{ij}^{\alpha})\log(1-p_{ij}^{\alpha})].
 \end{equation}
 If the number of constraints  is non extensive $M\times Q(Q+1)/2 \ll N$, this expression in the large $N$ limit is given by 
 \begin{eqnarray}
 S&=&-\sum_{\alpha=1}^M\sum_{q\neq q'} \log \left(\begin{array}{c} n_q n_{q'}\nonumber \\ e_{q,q'}^{\alpha}\end{array}\right)+\nonumber \\
 &&-\sum_{\alpha=1}^M\sum_q \log \left(\begin{array}{c}\frac{n_q (n_q-1)}{2}\nonumber \\ e_{q,q}^{\alpha}\end{array}\right).
 \end{eqnarray}
 
 \subsubsection{Multiplex ensemble with given expected degree sequence in each layer  and given expected number of links  present in each layer between nodes in different communities}
 We assign to each node $i$ a label $q_i=1,2\ldots, Q$ indicating the community to which node $i$ belongs.
 We can consider canonical uncorrelated multiplex ensembles in which we  fix the expected degree $k_i^{\alpha}$ of every node $i$ in each layer $\alpha$ together with the expected number of links $e_{q,q'}^{\alpha}$ between nodes in community $q$ and nodes in community $q'$ in layer $\alpha$.
 In this case we have $M\times N$ constraints  in the system indicated with a labels $\alpha=1,2,\ldots, M$ and $i=1,2\ldots, N$ and other $M\frac{Q(Q+1)}{2}$ constraints  indicated with labels $\alpha=1,2,\ldots, M$ and $q,q'=1,2\ldots,Q$. These constraints are given by  
 \begin{eqnarray}
 \sum_{\vec{G}} F_{i,\alpha}(\vec{G})P(\vec{G})&=&k_i^{\alpha} \label{4uca} \\
\sum_{\vec{G}} F_{q,q',\alpha}(\vec{G})P(\vec{G})&=&e_{q,q'}^{\alpha}\label{4ucb}
 \end{eqnarray}
 where the explicit expression for $F_{i,\alpha}(\vec{G})$ and for $F_{q,q',\alpha}(\vec{G})$ is given by
 \begin{eqnarray}
 F_{i,\alpha}(\vec{G})&=&\sum_{j=1,j\neq i}^Na_{ij}^{\alpha}\nonumber \\
 F_{q,q',\alpha}(G^{\alpha})&=&\sum_{i,j}a_{ij}^{\alpha}\delta_{q,q_i}\delta_{q',q_j}, \ \mbox{for} \  q\neq q' \nonumber \\
 F_{q,q,\alpha}(\vec{G})&=&\sum_{i<j}a_{ij}^{\alpha}\delta_{q,q_i}\delta_{q,q_j}.
 \end{eqnarray}
The probability of the multiplex is given by Eq. $(\ref{PC})$. Using this expression we observe that the probability $P_C(\vec{G})$ can be written as 
 \begin{eqnarray}
 P_C(\vec{G})&=&\frac{1}{Z_C}\exp\left[-\sum_{\alpha=1}^M\sum_{i=1}^N\lambda_{i,\alpha}F_{i,\alpha}(\vec{G})\right]\times\nonumber \\
 &&\times \exp\left[-\sum_{\alpha=1}^M\sum_{q\leq q'}\lambda_{q,q',\alpha}F_{q,q',\alpha}(\vec{G})\right]
 \label{4PCUL}
 \end{eqnarray}
 where $Z^{\alpha}_C$ is the normalization factor, $\lambda_{i,\alpha}$ is the Lagrangian multiplier enforcing the constraint  given by Eq. $(\ref{4uca})$ and $\lambda_{q,q',\alpha}$ is the Lagrangian multiplier enforcing the constraint given by Eq. $(\ref{4ucb})$.
 The probability of a link between node $i$ and node $j$ in layer $\alpha$ is given by 
 \begin{equation}
 p_{ij}^{\alpha}=\avg{a_{ij}^{\alpha}}=\frac{e^{-\lambda_{i,\alpha}-\lambda_{j,\alpha}-\lambda_{q,q',\alpha}}}{1+e^{-\lambda_{i,\alpha}-\lambda_{j,\alpha}-\lambda_{q,q',\alpha}}}
 \end{equation}
 where the Lagrangian multipliers are fixed by the conditions 
 \begin{eqnarray}
 \sum_{j=1,j\neq i}^N p_{ij}^{\alpha}=k_i^{\alpha}\nonumber \\
 \sum_{i,j} p_{ij}^{\alpha}\delta_{q,q_i}\delta_{q',q_j}&=&e_{q,q'}^{\alpha} \ \mbox{for}\ q\neq q'\nonumber \\
  \sum_{i<j} p_{ij}^{\alpha}\delta_{q,q_i}\delta_{q,q_j}&=&e_{q,q}^{\alpha}
 \end{eqnarray}
  Using the definition of the entropy of the multiplex Eq. $(\ref{entropy})$ and the expression for $P_C(\vec{G})$ given by Eq. $(\ref{4PCUL})$ it is easy to show that the entropy of the canonical multiplex ensemble $S$, that we call Shannon entropy, is given by 
 \begin{equation}
 S=-\sum_{\alpha}\sum_{i<j}[p_{ij}^{\alpha}\log p_{ij}^{\alpha}+(1-p_{ij}^{\alpha})\log(1-p_{ij}^{\alpha})].
 \label{Suc4}
 \end{equation}

 \subsection{Properties of  the uncorrelated canonical multiplex ensembles under consideration}
  In all the ensembles taken in consideration in the previous subsection the existence of any link is independent on the presence of other links in the multiplex and the probability of a given multiplex $\vec{G}$ is given by 
  \begin{equation}
  P_C(\vec{G})=\prod_{\alpha=1}^M\prod_{i<j} \left[p_{ij}^{\alpha}a_{ij}^{\alpha}+(1-p_{ij}^{\alpha})(1-a_{ij}^{\alpha})\right]
  \label{PG}
  \end{equation}
   Using the definition of the entropy of the multiplex Eq. $(\ref{entropy})$ and the expression for $P_C(\vec{G})$ given by Eq. $(\ref{PG})$ we can  show that the entropy of the canonical multiplex ensemble $S$, that we call Shannon entropy, is given by 
   \begin{equation}
 S=-\sum_{\alpha=1}^M\sum_{i<j}[p_{ij}^{\alpha}\log p_{ij}^{\alpha}+(1-p_{ij}^{\alpha})\log(1-p_{ij}^{\alpha})].
 \end{equation}
 for all the cases under consideration in subsection $\ref{eu}$.
 
  In the considered ensembles we can calculate  the average global overlap $\avg{O^{\alpha,\alpha'}}$ between two layers $\alpha$ and $\alpha'$ and the average local overlap $\avg{o^{\alpha,\alpha'}_i}$ between two layers $\alpha$ and $\alpha'$ where the global overlap $O^{\alpha,\alpha'}$ is defined in Eq. $(\ref{Og})$ and the local overlap $o^{\alpha,\alpha'}_i$ is defined in Eq. $(\ref{oi})$. These quantities are given by 
 \begin{eqnarray}
 \avg{O^{\alpha,\alpha'}}=\sum_{i<j}p^{\alpha}_{ij} p^{\alpha'}_{ij}\nonumber \\
 \avg{o_i^{\alpha,\alpha'}}=\sum_{j=1,j\neq i}^Np^{\alpha}_{ij} p^{\alpha'}_{ij}.
\label{AO} 
\end{eqnarray}
 
 For a multiplex ensemble with fixed expected total number of links $L^{\alpha}$ in each layer $\alpha$ we have $p^{\alpha}_{ij}=p^{\alpha}=2L^{\alpha}/[N(N-1)]$ and therefore,  
 \begin{eqnarray}
 \avg{O^{\alpha,\alpha'}}=\frac{2L^{\alpha}L^{\alpha'}}{N(N-1)}\nonumber \\
 \avg{o_i^{\alpha,\alpha'}}=\frac{4L^{\alpha}L^{\alpha'}}{N^2(N-1)}
 \end{eqnarray}
 Therefore if $L^{\alpha}={\cal O}(N)\ \forall \alpha=1,2,\ldots, M$, then the average global overlap is a finite number in the large network limit and the local overlap is vanishing in the large network limit. Therefore in this case the overlap of links is a totally negligible phenomena in the multiplex. In fact the average global overlap $\avg{O^{\alpha,\alpha'}}$ is much smaller than the total number of links in layer $\alpha$, $L^{\alpha}$ or the total number of links in layer $\alpha'$, i.e. $L^{\alpha'}$. Moreover the average local overlap $\avg{o_i^{\alpha,\alpha'}}$ is much smaller that the expected degree of node $i$ in  layer $\alpha$ or in layer $\alpha'$.
 For multiplex ensembles with given expected degree of the nodes in each layer, and with $k_i^{\alpha}<\sqrt{\avg{k^{\alpha}}N}$ we have  $p_{ij}^{\alpha}=\frac{k_i^{\alpha} k_j^{\alpha}}{\Avg{k^{\alpha}}N}$ and therefore
 \begin{eqnarray}
 \avg{O^{\alpha,\alpha'}}&=&\frac{1}{2} \left(\frac{\Avg{k^{\alpha}k^{\alpha'}}^2}{\Avg{k^{\alpha}}\Avg{k^{\alpha'}}}\right)\nonumber \\
 \avg{o_i^{\alpha,\alpha'}}&=&k_i^{\alpha}k_i^{\alpha'}\frac{\Avg{k^{\alpha}k^{\alpha'}}}{\Avg{k^{\alpha}}\Avg{k^{\alpha'}}N}
 \end{eqnarray}
 where $\Avg{k^{\alpha}k^{\alpha'}}=\sum_{i=1}^N k_i^{\alpha}k_i^{\alpha'}/N$.
 
 If the degrees in the different layers are uncorrelated (i.e. $\Avg{k^{\alpha}k^{\alpha'}}=\Avg{k^{\alpha}}\Avg{k^{\alpha'}}$) then the global and local overlaps are given by 
 \begin{eqnarray}
  \avg{O^{\alpha,\alpha'}}&=&\frac{1}{2}\left({\Avg{k^{\alpha}}\Avg{k^{\alpha'}}}\right)\ll N\nonumber \\
  \avg{o_i^{\alpha,\alpha'}}&=&\frac{k_i^{\alpha}k_i^{\alpha'}}{N}\ll \min(k_i^{\alpha},k_i^{\alpha'})
 \end{eqnarray}
 Therefore also in this case the overlap is negligible.
 Degree correlation in between different layers can enhance the overlap, but as long as $\Avg{k^{\alpha}k^{\alpha'}}\ll N$ the average global $ \avg{O^{\alpha,\alpha'}}$ and the local $\avg{o_i^{\alpha,\alpha'}}$ overlap continue to remain negligible with respect to the total number of nodes  in the two  layers and the degrees of the node $i$ in the two layers.
 Similarly using Eq. $(\ref{AO})$ it is possible to calculate the expected global overlap  and local overlap also in the multiplex ensemble in which we fix the number of links that in layer connect nodes belonging to different communities   and in the multiplex ensemble in which we fix at the same time the average degree of each node in each layer and the average number of links in between nodes of different communities at any given layer.
In general if in a multiplex ensemble we want to have a given  significant overlap we need to consider correlated multiplex ensembles.
 \subsection{Construction of a uncorrelated multiplex in an uncorrelated canonical multiplex ensemble under consideration}
 In all the cases taken into consideration in the previous subsections, the 
 probability of a network $G^{\alpha}$ on layer $\alpha$ is uncorrelated with the other networks in the other layers. In particular,  the probability of a multiplex $\vec{G}$ can  be written as in Eq. $(\ref{PG})$.

 Therefore in order to construct a multiplex in the canonical network ensembles it is sufficient to follow the following scheme
\begin{itemize}
\item
Calculate the probability $p_{ij}^{\alpha}$ to have a link between node $i$ and $j$ in layer $\alpha$.
\item
For every pair of node $i$ and $j$ put a link in layer $\alpha$ with probability $p_{ij}^{\alpha}$. Do this for every layer $\alpha=1,2,\ldots, M $ independently.
\end{itemize}

\subsection{Examples of correlated canonical multiplex ensembles}
\label{ec}

If the probability of a multiplex $P_C(\vec{G})$ does not factorize into the probabilities $P_C^{\alpha}({G}^{\alpha})$ of the networks in the different layers $\alpha$ of the multiplex, i.e. 
if 
\begin{equation}
P_C(\vec{G})\neq \prod_{\alpha=1}^M P_C^{\alpha}(G^{\alpha})
\end{equation} the multiplex is correlated.
In these ensembles the existence of a link in one layer can be correlated with the existence of a link in another layer.
For single networks, when we want to treat ensembles in which the links are correlated we need to make use of a parametrization that takes into account not only of single independent links but also of correlated set of links called   subgraphs, such a triangles, triples, and so on \cite{Ne1,NeK2}. Similarly if we want to treat correlated multiplex, it is convenient to consider multilinks. In this way our multiplex is not anymore described by $M$ adjacency matrices describing the networks at each multiplex layer, but the network is described by a much larger set of variables corresponding to correlated links, i.e. multilinks, and is fully characterized by 
$2^M$ multiadjacency matrices.
The simplest case of correlated multiplex ensemble is an ensemble in which we fix the expected total number of multilinks $\vec{m}$ in the network defined in section $ \ref{ml}$.
Starting from this example of correlated canonical multiplex ensemble we can generate more refined models in which we fix the expected multidegree sequence $k_i^{\vec{m}}$ defined in section $\ref{ml}$ or the expected number of multilinks  $\vec{m}$ linking nodes of different communities etc.
In the following we will describe in detail some of the more relevant examples of correlated canonical multiplex ensembles.

\subsubsection{Multiplex ensemble with given expected total number of multilinks $\vec{m}$}
\label{ec1}

 We can fix the average number $L^{\vec{m}}$ of multilinks  $\vec{m}$ with the condition $\sum_{\vec{m}}L^{\vec{m}}=N(N-1)/2$.
 In this case we have $K=2^M$ constraints  indicated by the label $\vec{m}=(m_1,m_2,\ldots,m_{\alpha},\ldots,m_M)$ with $m_{\alpha}=0,1$. These constraints are given by 
 \begin{equation}
\sum_{\vec{G}} F_{\vec{m}}(\vec{G})P_C(\vec{G})=\sum_{\vec{G}}\sum_{i<j}A_{ij}^{\vec{m}}P_C(\vec{G})={L^{\vec{m}}}
\label{1ucc}
 \end{equation}
where the multiadjacency matrices of elements $A_{ij}^{\vec{m}}$ are defined 
 in Eq. ($\ref{MA}$).
 In this case the functions $F_{\vec{m}}(\vec{G})$ are given by 
 \begin{equation}
 F_{\vec{m}}(\vec{G})=\sum_{i<j}A_{ij}^{\vec{m}}.
 \end{equation}
 The probability $P_C(\vec{G})$ of a multiplex in the ensemble  is given by Eq. $(\ref{PC})$ that reads in this specific case 
 \begin{equation}
 P_C({\vec{G}})=\frac{1}{Z_C}\exp\left[-\sum_{\vec{m}}\lambda_{\vec{m}}\sum_{i<j}A_{ij}^{\vec{m}}\right]
 \label{1PCC}
 \end{equation}
  where $Z_C$ is the canonical partition function and $\lambda_{\vec{m}}$ is the Lagrangian multiplier enforcing the constraint given by Eq. $(\ref{1ucc})$.
 The probability $p_{ij}^{\vec{m}}$ of a multilink $\vec{m}$ between node $i$ and node $j$ is given by 
 \begin{equation}
 p_{ij}^{\vec{m}}=p^{\vec{m}}=\Avg{A_{ij}^{\vec{m}}}=\frac{e^{-\lambda_{\vec{m}}}}{\sum_{\vec{m}}e^{-\lambda_{\vec{m}}}}
 \label{pm}
 \end{equation}
 with $\sum_{i<j}p_{ij}^{\vec{m}}=L^{\vec{m}}$, and $\sum_{\vec{m}}p_{ij}^{\vec{m}}=1$ implying
 \begin{equation}
 p^{\vec{m}}=\frac{{L^{\vec{m}}}}{N(N-1)/2}.
 \label{pmcc1}
 \end{equation}
 The entropy of the canonical multiplex ensemble $S$ given by Eq.($\ref{entropy}$) can be calculated using the expression for $P_C(\vec{G})$ Eq. $(\ref{1PCC})$, obtaining
 \begin{equation}
 S=-\frac{N(N-1)}{2}\sum_{\vec{m}}[p^{\vec{m}}\log p^{\vec{m}}]
 \end{equation}
 with $p^{\vec{m}}$ is given by Eq. $(\ref{pmcc1})$.
 If the number of layers $M$ is finite this entropy $S$  is given by 
 \begin{equation}
 S=\log \left[\frac{\left(\frac{N(N-1)}{2}\right)!}{\prod_{\vec{m}}(L^{\vec{m}}!)}\right].
 \end{equation}

 \subsubsection{Multiplex ensemble with given  expected multidegree sequence}
 \label{ec2}
  We can fix the average  multidegree $k_i^{\vec{m}}$  of node  $i$ with the condition $\sum_{\vec{m}}k_i^{\vec{m}}=N-1$.  In this case we have $K=2^M\times N$ constraints  indicated by the label $\vec{m}=(m_1,m_2,\ldots,m_{\alpha},\ldots,m_M)$ with $m_{\alpha}=0,1$ and the label  $i=1,2,\ldots, N$. In particular we have,
 \begin{equation}
 \sum_{\vec{G}} F_{i,\vec{m}}(\vec{G})P_C(\vec{G})=\sum_{\vec{G}}\sum_{j}A_{ij}^{\vec{m}}P_C(\vec{G})=k_i^{\vec{m}}
 \label{2CC}
 \end{equation}
 for all $\vec{m}$ with $m_{\alpha}=0,1$ and all $i=1,2,\ldots N$, where
 the multiadjacency matrices of elements $A_{ij}^{\vec{m}}=0,1$ are given by 
Eq. $(\ref{MA})$.
 Therefore the functions  $F_{i,\vec{m}}(\vec{G})$ are given in this case by 
 \begin{equation}
 F_{i,\vec{m}}(\vec{G})=\sum_{j=1,j\neq i}^NA_{ij}^{\vec{m}}.
 \end{equation}
 The probability of the multiplex is given by Eq. $(\ref{PC})$ that in this case reads
 \begin{equation}
 P({\vec{G}})=\frac{1}{Z_C}\exp\left[-\sum_{\vec{m}}\sum_{i=1}^N\lambda_{i,\vec{m}}\sum_{j=1}^N A_{ij}^{\vec{m}}\right]
 \label{2PCC}
 \end{equation}
   where $Z_C$ is the canonical partition function and $\lambda_{i,\vec{m}}$ is the Lagrangian multiplier enforcing the constraint given by Eq. $ (\ref{2CC})$.
 The probability of a multilink $\vec{m}$ between node $i$ and node $j$ is given by 
 \begin{equation}
 p_{ij}^{\vec{m}}=\Avg{A_{ij}^{\vec{m}}}=\frac{e^{-\lambda_{i,\vec{m}}-\lambda_{j,\vec{m}}}}{\sum_{\vec{m}} e^{-\lambda_{i,\vec{m}}-\lambda_{j,\vec{m}}}}
 \end{equation}
 with  the Lagrangian multipliers $\lambda_{i,\vec{m}}$ fixed by the constraints

 \begin{eqnarray} 
 \sum_{\vec{m}}p_{ij}^{\vec{m}}&=&1\nonumber \\
 \sum_{j=1}^N p_{ij}^{\vec{m}}&=&k_i^{\vec{m}}.
 \end{eqnarray}The entropy of the canonical multiplex ensemble $S$, that we call Shannon entropy, is  given by Eq.($\ref{entropy}$) and  can be calculated using the expression for $P_C(\vec{G})$ Eq. $(\ref{2PCC})$, obtaining
 \begin{equation}
 S=-\sum_{\vec{m}}\sum_{i<j}[p_{ij}^{\vec{m}}\log p_{ij}^{\vec{m}}].
 \label{Skim}
 \end{equation}
 If the multiplex is sparse, i.e.  $k_i^{\vec{m}}<\sqrt{\avg{k^{\vec{m}}}N}$ provided that in the multilink $\vec{m}$ there is at least a link, i.e. $\sum_{\alpha=1}^M m_{\alpha}>0$, we have 
 \begin{equation}
 p_{ij}^{\vec{m}}=\frac{k_i^{\vec{m}}k_j^{\vec{m}}}{\avg{k^{\vec{m}}}N}
 \end{equation}
 for all $\vec{m}$ such that $\sum_{\alpha=1}^Mm_{\alpha}>0$.
  In this limit the entropy $S$ is given by 
 \begin{eqnarray}
 S&\simeq&\sum_{\vec{m}|\sum_{\alpha=1}^M m_{\alpha}>0}\left[-\sum_i k_i^{\vec{m}}\log(k_i^{\vec{m}})+\frac{1}{2}\avg{k^{\vec{m}}}N\right.\nonumber \\
&& \left.+\frac{1}{2}\avg{k^{\vec{m}}}N\log(\avg{k^{\vec{m}}}N)-\frac{1}{4}\left(\frac{\avg{(k^{\vec{m}})^2}}{\avg{k^{\vec{m}}}}\right)^2\right].
\label{Skimu}
 \end{eqnarray}
\subsubsection{Multiplex ensemble with given expected number of multilinks $\vec{m}$ between nodes in different communities}
We can fix the expected number of multilinks $\vec{m}$  between nodes in different communities of the multiplex.
 We assign to each node $i$ a discrete variable $q_i=1,2,\ldots, Q$ indicating the community of the node. 
 
  We can consider canonical uncorrelated multiplex ensembles in which we  fix  the expected number of multilinks $\vec{m}$, $e_{q,q'}^{\vec{m}}$ between nodes in community $q$ and nodes in community $q'$. Moreover we choose $e_{q,q'}^{\vec{m}}$ such that they satisfy the condition that the sum over the different multilinks $\vec{m}$ of $e_{q,q'}^{\vec{m}}$ is equal to the total number of links in between nodes in community $q$ and nodes in community $q'$.
 In this case we have $K=2^M\times Q(Q+1)/2$ constraints  in the system indicated with a labels $\vec{m}=(m_1,m_2,\ldots, m_M$ with $m_{\alpha}=0,1$ and the labels $q,q'=1,2,\ldots Q$. These constraints are given by 
 \begin{eqnarray}
\sum_{\vec{G}} F_{q,q',\vec{m}}(\vec{G})P_C(\vec{G})&=&e_{q,q'}^{\vec{m}}
\label{3CC}
 \end{eqnarray}
where  the explicit expression for $F_{q,q',\vec{m}}(\vec{G})$ is given by
 \begin{eqnarray}
 F_{q,q',\vec{m}}(\vec{G})&=&\sum_{i,j}A_{ij}^{\vec{m}}\delta_{q,q_i}\delta_{q',q_j}, \ \mbox{for} \  q\neq q' \nonumber \\
 F_{q,q,\vec{m}}(\vec{G})&=&\sum_{i<j}A_{ij}^{\vec{m}}\delta_{q,q_i}\delta_{q,q_j}, \end{eqnarray}
 and the multiadjacency matrices of elements $A_{ij}^{\vec{m}}$ are defined in Eq. $(\ref{MA})$.
 The probability of the multiplex is given by Eq. $(\ref{PC})$ and in this specific case is given by 
 \begin{equation}
 P_C(\vec{G})=\frac{1}{Z_C}\exp\left[-\sum_{\vec{m}}\sum_{q\leq q'}\lambda_{q,q',\vec{m}} F_{q,q',\vec{m}}(\vec{G})\right]
 \label{PCcc}
 \end{equation}
 where $Z_C$ is the canonical partition function and $\lambda_{q,q',\vec{m}}$ is the Lagrangian multiplier enforcing the constraint  given by Eq. $(\ref{3CC})$.
 The probability of a multilink $\vec{m}$ between node $i$ and node $j$  is given by 
 \begin{equation}
 p_{ij}^{\vec{m}}=\avg{A_{ij}^{\vec{m}}}=\frac{e^{-\lambda_{q_i,q_j,\vec{m}}}}{\sum_{\vec{m}}e^{-\lambda_{q_i,q_j,\vec{m}}}}
 \label{pijbc}
 \end{equation}
 where the Lagrangian multipliers are fixed by the conditions 
 \begin{eqnarray}
  \sum_{\vec{m}}p_{ij}^{\vec{m}}=1\nonumber \\
 \sum_{i,j} p_{ij}^{\vec{m}}\delta_{q,q_i}\delta_{q',q_j}&=&e_{q,q'}^{\vec{m}} \ \mbox{for}\ q\neq q'\nonumber \\
  \sum_{i<j} p_{ij}^{\vec{m}}\delta_{q,q_i}\delta_{q,q_j}&=&e_{q,q}^{\vec{m}}.
 \end{eqnarray}
As it can be seen by Eq. $(\ref{pijbc})$ the probabilities $p_{ij}^{\vec{m}}$ depend only on $q_i, q_j$ and $\vec{m}$ therefore we have $ p_{ij}^{\vec{m}}=p^{\vec{m}}(q_i,q_j)$ with

\begin{eqnarray}
p^{\vec{m}}(q,q')&=&\frac{e_{q,q'}^{\vec{m}}}{n_q n_{q'}} \ \mbox{for} \ q\neq q' \nonumber \\
p^{\vec{m}}(q,q)&=& \frac{e_{q,q}^{\vec{m}}}{n_q(n_q-1)/2}
\end{eqnarray}
where $n_q$ indicates the total number of nodes in community $q$.
 The entropy of the canonical multiplex ensemble $S$ that we call Shannon entropy is given by Eq. $(\ref{entropy})$. Evaluating this expression using the probability of the multiplex $P_C(\vec{G})$ given by $(\ref{PCcc})$ we obtain, 
 \begin{equation}
 S=-\sum_{\vec{m}}\sum_{i<j}[p_{ij}^{\vec{m}}\log p_{ij}^{\vec{m}}].
 \end{equation}
 If the number of constraints  is non extensive $2^M Q(Q+1)/2 \ll N$, this expression in the large $N$ limit is given by 
 \begin{eqnarray}
 S&=&\sum_{q\neq q'} \log \left[\frac{(n_q n_{q'})!}{\prod_{\vec{m}} (e_{q,q'}^{\vec{m}}!)}\right]+\nonumber \\
 &&+\sum_q \log \left[\frac{\left(\frac{n_q (n_q-1)}{2}\right)!}{ \prod_{\vec{m}}(e_{q,q}^{\vec{m}}!)}\right].
 \end{eqnarray}

\subsubsection{Multiplex ensemble with  fixed expected multidegree sequence and expected number of multilinks $\vec{m}$ between nodes in different communities}
 We assign to each node $i$ a label $q_i=1,2\ldots, Q$ indicating the community to which node $i$ belongs.
 We can consider canonical uncorrelated multiplex ensembles in which we  fix the expected multidegree $k_i^{\vec{m}}$ of every node $i$  (with the condition $\sum_{\vec{m}}k_i^{\vec{m}}=N-1$) together with the expected number of multilinks $e_{q,q'}^{\vec{m}}$ between nodes in community $q$ and nodes in community $q'$  (with the condition that the sum over the different multilinks $\vec{m}$ of $e_{q,q'}^{\vec{m}}$ is equal to the total number of links in between nodes in community $q$ and nodes in community $q'$).
 In this case we have $2^M\times N$ constraints   indicated with a labels $\vec{m}=(m_1,m_2,\ldots, m_M)$ with $m_{\alpha}=0,1$ and $i=1,2\ldots, N$ and other $2^M\times \frac{Q(Q+1)}{2}$ constraints  indicated with labels $\vec{m}$ and $q,q'=1,2\ldots,Q$. These constraints are given by 
 \begin{eqnarray}
 \sum_{\vec{G}} F_{i,\vec{m}}(\vec{G})P(\vec{G})&=&k_i^{\vec{m}}\label{4cca}\\
\sum_{\vec{G}} F_{q,q',\vec{m}}P_C(\vec{G})&=&e_{q,q'}^{\vec{m}} \ \mbox{for} \ q\neq q' \label{4ccb}\\
\sum_{\vec{G}}F_{q,q,\vec{m}}(\vec{G})P_C(\vec{G})&=&e_{q,q'}^{\vec{m}}\label{4ccc}
 \end{eqnarray}
 where the explicit expression for $F_{i,\vec{m}}(\vec{G})$ and for $F_{q,q',\vec{m}}(\vec{G})$ are given by
 \begin{eqnarray}
 F_{i,\vec{m}}(\vec{G})&=&\sum_{j}A_{ij}^{\vec{m}}\nonumber \\
 F_{q,q',\vec{m}}(\vec{G})&=&\sum_{i,j}A_{ij}^{\vec{m}}\delta_{q,q_i}\delta_{q',q_j}, \ \mbox{for} \  q\neq q' \nonumber \\
 F_{q,q,\vec{m}}(\vec{G})&=&\sum_{i<j}A_{ij}^{\vec{m}}\delta_{q,q_i}\delta_{q,q_j},
 \end{eqnarray}
 where the element $A_{ij}^{\vec{m}}$ of the multiadjacency matrices is defined in Eq. $(\ref{MA})$.
 The probability of the multiplex is given by Eq. $(\ref{PC})$ that reads in this case 
 \begin{eqnarray}
 P_C(\vec{G})&=&\frac{1}{Z_C}\exp\left[-\sum_{\vec{m}}\sum_{i}\lambda_{i,\vec{m}}F_{i,\vec{m}}(\vec{G})\right]\times \nonumber \\
 &\times &\exp\left[-\sum_{\vec{m}}\sum_{q\leq q'}\lambda_{q,q'\vec{m}}F_{q,q'\vec{m}}(\vec{G})\right]
 \end{eqnarray}
 where $Z_C$ is the canonical partition function, $\lambda_{i,\vec{m}}$ is the Lagrangian multiplier enforcing the constraint  given by Eq. $(\ref{4cca})$ and $\lambda_{q,q',\vec{m}}$ is the Lagrangian multiplier enforcing the constraint given by Eq. $(\ref{4ccb})$ or by Eq. $(\ref{4ccc})$.
 The probability of a multilink  $\vec{m}$ between node $i$ and node $j$  is given by 
 \begin{equation}
 p_{ij}^{\vec{m}}=\Avg{A_{ij}^{\vec{m}}}=\frac{e^{-\lambda_{i,\vec{m}}-\lambda_{j,\vec{m}}-\lambda_{q,q',\vec{m}}}}{\sum_{\vec{m}}e^{-\lambda_{i,\vec{m}}-\lambda_{j,\vec{m}}-\lambda_{q,q',\vec{m}}}}
 \label{lck}
 \end{equation}
 where the Lagrangian multipliers are fixed by the conditions 
 \begin{eqnarray}
 \sum_{\vec{m}}p_{ij}^{\vec{m}}=1\nonumber \\
 \sum_j p_{ij}^{\vec{m}}=k_i^{\vec{m}}\nonumber \\
 \sum_{i,j} p_{ij}^{\vec{m}}\delta_{q,q_i}\delta_{q',q_j}&=&e_{q,q'}^{\vec{m}} \ \mbox{for}\ q\neq q'\nonumber \\
  \sum_{i<j} p_{ij}^{\vec{m}}\delta_{q,q_i}\delta_{q,q_j}&=&e_{q,q}^{\vec{m}}
  \label{cck}
 \end{eqnarray}
 The entropy of the canonical multiplex ensemble that we call Shannon entropy is given by 
 \begin{equation}
 S=-\sum_{\vec{m}}\sum_{i<j}[p_{ij}^{\vec{m}}\log p_{ij}^{\vec{m}}].
 \label{Sck}
 \end{equation}
 where the probabilities $p_{ij}^{\vec{m}}$ are given by Eq. $(\ref{lck})$ and satisfy Eqs. ($\ref{cck}$).
\subsection{Overlap in correlated canonical ensembles under consideration}
 In all the cases taken into consideration in the previous subsection, the 
 probability of a network $G^{\alpha}$ on layer $\alpha$ is correlated with the other networks in the other layers. Therefore the probability $P_C(\vec{G})$ cannot be factorized in the probability for single layers.
 Nevertheless $P_C(\vec{G})$ takes a simple form in the cases that we have investigated so far, i.e. 
 \begin{equation}
 P_C(\vec{G})=\prod_{i<j}\left[\prod_{\vec{m}} p_{ij}^{\vec{m}}A_{ij}^{\vec{m}}\right].
 \label{PCC}
 \end{equation}
 where $\vec{m}=(m_1,m_2,\ldots, m_M)$ is a vector  of elements $m_{\alpha}=0,1$ and where 
$ A_{ij}^{\vec{m}}$ are the  multiadjacency matrices defined in Eq. $(\ref{MA})$.
In these ensembles the Shannon entropy $S$ given by Eq. $(\ref{entropy})$
takes the simple form
\begin{equation}
S=-\sum_{i<j}\sum_{\vec{m}} p_{ij}^{\vec{m}}\log p_{ij}^{\vec{m}}.
\label{entropym}
\end{equation}
 In the considered ensembles we can calculate  the average total overlap $\avg{O^{\alpha,\alpha'}}$ between two layers $\alpha$ and $\alpha'$ and the average local overlap $\avg{o^{\alpha,\alpha'}_i}$ between two layers $\alpha$ and $\alpha'$, where the global overlap $O^{\alpha,\alpha'}$ is defined in Eq. $(\ref{Og})$ and the local overlap $o^{\alpha,\alpha'}_i$ is defined in Eq. $(\ref{oi})$. These quantities are given by 
 \begin{eqnarray}
 \avg{O^{\alpha,\alpha'}}=\sum_{\vec{m}|m_{\alpha}=1, m_{\alpha'}=1}\sum_{i<j}p^{\vec{m}}_{ij}\nonumber \\
 \avg{o_i^{\alpha,\alpha'}}=\sum_{\vec{m}|m_{\alpha}=1, m_{\alpha'}=1}\sum_{j=1}^Np^{\vec{m}}_{ij}.
\label{AOcorr} 
\end{eqnarray}
These quantities now can be significant also for sparse networks as we will see in the next subsection in the simple case of a multiplex with just two layers, i.e.  $M=2$. 
\subsection{Case of a two layers multiplex, i.e.  $M=2$}
Let us  consider the simple case of a correlated  multiplex ensembles formed by   $M=2$ layers, network 1 and network 2.
The probability   $P_C(\vec{G})$ of a multiplex  in all the cases taken in consideration in the  subsection $\ref{ec}$, is given by Eq. $(\ref{PCC})$ that reads in this case  
\begin{eqnarray}
P(\vec{G})&=&\prod_{i<j} \left[p^{00}_{ij}(1-a^1_{ij})(1-a^2_{ij})+p^{10}_{ij}a^1_{ij}(1-a^2_{ij})\right. \nonumber \\ 
&& \left.+p^{01}_{ij}{(1-a^1_{ij})a^2_{ij}}+p^{11}_{ij} {a_{ij}^1 a^2_{ij}}\right]
\end{eqnarray} 
where $p^{n_1,n_2}_{ij}$ is the probability to have $n_1=0,1$ links between node $i$ and node $j$ in network $1$ and $n_2=0,1$ links between the same nodes in network $2$. The probabilities $p_{ij}^{n_1 n_2}$ satisfy the constrain $p_{ij}^{00}+p^{01}_{ij}+p^{10}_{ij}+p^{11}_{ij}=1$.
The entropy of such multiplex is then given by Eq. $(\ref{entropym})$ that reads in this case
\begin{equation}
S=-\sum_{n_1,n_2}\sum_{i<j}p^{n_1n_2}_{ij}\ln {p^{n_1n_2}_{ij}}.
\end{equation}

In the considered ensembles we can calculate  the average total overlap $\avg{O^{1,2}}=\avg{O}$ between two layers $1$ and $2$ and the average local overlap $\avg{o^{1,2}_i}=\avg{o_i}$  defined in Eqs. $(\ref{AOcorr})$. 
For the ensembles in which we fix the expected total number of multilinks $\vec{m}$,
$L^{\vec{m}}$ considered in subsection $\ref{ec1}$ we have 
 \begin{eqnarray}
 \avg{O}&=&L^{{11}}\nonumber \\
 \avg{o_i}&=&\frac{2L^{11}}{N-1}. 
 \label{Ooc}
\end{eqnarray}
Assuming $L^{11},L^{10}, L^{01}\propto N$, Eq. $(\ref{Ooc})$ implies that the fraction of links that overlap is not negligible (globally and locally) also if both network 1 and network 2 are sparse.
For the ensemble in which we fix the expected  multidegree (considered in subsection $\ref{ec2}$), considering the additional condition $k_i^{\vec{m}}<\sqrt{\avg{k^{\vec{m}}}N}$ for all multilinks $\vec{m}$ formed at least by a link, i.e. $\sum_{\alpha=1}^M m_{\alpha}>0$, we have   $p_{ij}^{\vec{m}}=\frac{k_i^{\vec{m}}k_j^{\vec{m}}}{\avg{k^{\vec{m}}}N}$ and therefore, 
 \begin{eqnarray}
 \avg{O}&=&\frac{1}{2}\avg{k^{11}}N\nonumber \\
 \avg{o_i}&=&k_i^{11}. 
\end{eqnarray}
Provided that $\avg{k^{11}}$ is finite,  we find  that also in this case the global and local overlap can be significant also if both network 1 and network 2 are sparse.
A similar conclusion can be drawn for the other two cases of correlated multiplex ensembles taken in consideration in the previous paragraphs.

\subsection{Construction of correlated multiplex in the canonical multiplex ensemble}

 Since in the considered cases of correlated multiplex ensemble the probability of a multiplex can be expressed as in Eq. $(\ref{PCC})$, in order  to construct a correlated multiplex in the canonical network ensembles it is sufficient to follow the following scheme.
\begin{itemize}
\item
Calculate the probability $p_{ij}^{\vec{m}}$ to have a multilink $\vec{m}$ between node $i$ and $j$.
\item
For every pair of node $i$ and $j$,  draw a multilink $\vec{m}$ with probability $p_{ij}^{\vec{m}}$ and consequently  put a  link in every  layer $\alpha$ where $m_{\alpha}=1$ and put no link in every layer $\alpha$ where $m_{\alpha}=0$.
\end{itemize}

\section{Microcanonical multiplex  ensembles}

The microcanonical multiplex ensembles are formed by the multiplexes that satisfy some hard constraints. Every multiplex  in a microcanonical multiplex ensemble has equal probability. We note here that we consider only graphical constraints \cite{DelGenio1}, i.e. constraints that can be satisfied at least by one realization of the multiplex. This is a condition that for example is automatically satisfied  if we consider network ensembles that are a randomization of a real multiplex with some given structural features.
Therefore the probability $P_M(\vec{G})$ of a microcanonical multiplex ensemble is given by 
\begin{equation}
P_M(\vec{G})=\frac{1}{Z_M}\prod_{\mu=1}^P \delta[F_{\mu}(\vec{G}), C_{\mu}]
\label{PM}
\end{equation}
where $\delta[]$ is the Kronecker delta and where $Z_M$ is the ``microcanonical partition function" of the multiplex given by  
\begin{equation}
Z_M=\sum_{\vec{G}}\prod_{\mu=1}^P \delta[F_{\mu}(\vec{G}), C_{\mu}].
\label{ZM}
\end{equation}
Therefore the microcanonical partition function $Z_M$ of the multiplex ensemble counts the number of multiplexes satisfying the hard constraints $F_{\mu}(\vec{G})=C_{\mu}$ for  $\mu=1,2\ldots,P$.
We call the entropy of these multiplex ensembles  $N\Sigma$ and using the definition of the entropy of an ensemble given by Eq.$(\ref{entropy})$ together with the expression for the probability of a multiplex in the microcanonical ensemble given by Eq. $(\ref{PM})$ we have
\begin{equation}
N\Sigma=-\sum_{\vec{G}}P_M(\vec{G})\log P_M(\vec{G})=\log Z_M,
\label{SM}
\end{equation}
where we call $\Sigma$ the {\it Gibbs entropy} of the multiplex ensemble.
The Gibbs entropy $\Sigma$ of microcanonical multiplex  ensembles  is related to the Shannon entropy $S$ of the associated canonical multiplex ensemble $S$ which enforce the same constraint of the microcanonical network ensemble in average (the conjugated canonical ensemble), by a simple relation. In fact we have
\begin{equation}
N\Sigma=S-N\Omega
\label{SSO}
\end{equation}
where $N\Omega$ is  equal to the logarithm of the probability that in the conjugated canonical multiplex ensemble the hard constraints $F_{\mu}(\vec{G})$ are satisfied, i.e.
\begin{equation}
N\Omega=-\log\left\{\sum_{\vec{G}}P_C(\vec{G})\prod_{\mu=1}^M\delta[F_{\mu}(\vec{G}),C_{\mu}]\right\}.
\label{Omega}
\end{equation}
In order to verify the relation Eq. $(\ref{SSO})$ we observe that the canonical multiplex probability $P_C(\vec{G})$ is given by Eq. $(\ref{PC})$ that we rewrite here for convenience, 
\begin{equation}
P_C(\vec{G})=\frac{1}{Z_C}e^{-\sum_{\mu=1}^P\lambda_{\mu} F_{\mu}(\vec{G})}
\end{equation}
and therefore, using Eq. $(\ref{Omega})$  we get
\begin{eqnarray}
\exp[-N\Omega]&=&\sum_{\vec{G}} \frac{1}{Z_C}e^{-\sum_{\mu=1}^P\lambda_{\mu} F_{\mu}(\vec{G})}\prod_{\mu=1}^P\delta[F_{\mu}(\vec{G}),C_{\mu}]\nonumber \\
&=&\frac{1}{Z_C}e^{-\sum_{\mu=1}^P \lambda_{\mu}C_{\mu}}\sum_{\vec{G}}\prod_{\mu=1}^P\delta[F_{\mu}(\vec{G}),C_{\mu}]\nonumber \\
&=& \frac{Z_M}{e^{S}}=\exp[N\Sigma-S].
\end{eqnarray}
where in the last relation we have used Eq. $(\ref{Sc})$, Eq. $(\ref{ZM})$ and Eq. $(\ref{SM})$.
Given Eq. $(\ref{SSO})$, if $\Omega$ is larger than zero in the limit $N\gg 1$, the microcanonical and the conjugated canonical multiplex ensemble are not equivalent.

\subsection{Uncorrelated  microcanonical multiplex ensembles}

In an uncorrelated multiplex ensemble we have that the probability of a multiplex $\vec{G}$ is factorizable into the product of probabilities $P_{\alpha}(G^{\alpha})$ of the networks $G^{\alpha}$ in layer $\alpha$, i.e.
\begin{equation}
P_M(\vec{G})=\prod_{\alpha=1}^M P_M^{\alpha}(G^{\alpha}).
\label{UC2}
\end{equation}
Given the general expression for $P_M(\vec{G})$ provided by Eq. $(\ref{PM})$ we can conclude that a microcanonical multiplex ensemble is uncorrelated only if  the hard constraints $F_{\mu}(\vec{G})=C_{\mu}$ with $\mu=1,2\ldots, K$  involve for every constraint $\mu$ only one network $G^{\alpha}$ in one layer $\alpha$ of the multiplex.
Therefore we will indicate the function $F_{\mu}(\vec{G})$ with a label indicating the layer $\alpha$ and one label $\nu$ counting the number of constraints in each layer, i.e. $F_{\nu,\alpha}(G^{\alpha})$.

Given the condition Eq. $(\ref{UC2})$ the Gibbs entropy $\Sigma$ of the multiplex 
can be expressed as in the following
\begin{equation}
N\Sigma=-\sum_{\vec{G}}P_M(\vec{G})\log P_M(\vec{G})=\sum_{\alpha} N\Sigma^{\alpha}.
\end{equation}
 where $\Sigma^{\alpha}$ is the Gibbs entropy of the network ensemble induced in layer $\alpha$,
 \begin{equation}
 N\Sigma^{\alpha}=-\sum_{G^{\alpha}} P_M^{\alpha}(G^{\alpha})\log P_M^{\alpha}(G^{\alpha})
 \end{equation}
 with $P_M^{\alpha}(G^{\alpha})=\prod_{\nu}\delta[F_{\nu,\alpha}(G^{\alpha}),C_{\nu,\alpha}]/Z_M^{\alpha}$ and 
 \begin{equation}
 Z_{M}^{\alpha}=\sum_{G^{\alpha}} \prod_{\nu}\delta[F_{\nu,\alpha}(G^{\alpha}),C_{\nu,\alpha}].
 \end{equation}
 Using the same arguments used to derive Eq. $(\ref{SSO})$ it is straightforward to show that the Gibbs entropy $\Sigma^{\alpha}$ of each network ensemble  at layer $\alpha$ is given by
 \begin{equation}
 N\Sigma^{\alpha}=S^{\alpha}-N\Omega^{\alpha}
 \label{rsso}
 \end{equation}
 where $S^{\alpha}$ is the Shannon entropy of the canonical network ensemble which enforce the same constraint of the microcanonical network ensemble in average, i.e.
 \begin{equation}
 S^{\alpha}=-\sum_{G^{\alpha}}P_C^{\alpha}(G^{\alpha})\log P_C^{\alpha}(G^{\alpha}),
 \end{equation}
 where   $P_C^{\alpha}(G^{\alpha})$ is the  probability for a network $G^{\alpha}$ in layer $\alpha$.
 Moreover  $\Omega^{\alpha}$ in Eq. $(\ref{rsso})$ satisfies
\begin{equation}
N\Omega^{\alpha}=-\log \left\{\sum_{G^{\alpha}} P_C^{\alpha}({G^{\alpha}})\prod_{\nu}\delta[F_{\nu,\alpha}(G^{\alpha}),C_{\nu,\alpha}]\right\}.
\end{equation}
Examples of uncorrelated microcanonical multiplex ensemble are given by ensembles in which we fix the total number of links at each layer, the degree sequence at each layer, the number of links between nodes in different communities in each layer etc. In the following subsection we present in detail several examples of uncorrelated microcanonical multiplex ensembles.

\subsection{Examples of uncorrelated microcanonical multiplex ensembles}
\subsubsection{Multiplex ensemble with given  total number of links in each layer}
 We can fix the total  number of links $L^{\alpha}$ in each layer $\alpha$ of the multiplex. In this case we have $K=M$ constraints  in the system indicated with a label $\alpha=1,2,\ldots, M$. These constraints are given by 
 \begin{equation}
F_{\alpha}(\vec{G})=L^{\alpha}
 \end{equation}
 with $\alpha=1,2,\ldots, M$ and with $F_{\alpha}(\vec{G})$  given by
 \begin{equation}
 F_{\alpha}(\vec{G})=\sum_{i<j}a_{ij}^{\alpha}.
 \end{equation}
The microcanonical partition function $Z_M$ is equal to the number of multiplexes in these ensemble, which is given by the product over the layers $\alpha=1,2\ldots, M$ of the number of networks $G^{\alpha}$ satisfying the constraints $F_{\alpha}(\vec{G})=L^{\alpha}$. The number of networks $G^{\alpha}$ with $L^{\alpha}$ links is given by the number of ways of choosing $L^{\alpha}$ links out of $N(N-1)/2$ possible links, we have therefore
\begin{equation}
Z_M=\prod_{\alpha=1}^M \left(\begin{array}{c}\left(\begin{array}{c} N\\2 \end{array}\right) \\L^{\alpha}\end{array}\right).
\end{equation} 
Using Eq. $(\ref{SM})$ we find that  the Gibbs entropy for this ensemble is given by 
\begin{equation}
N\Sigma=\log \left(\begin{array}{c}\left(\begin{array}{c} N\\2 \end{array}\right) \\L^{\alpha}\end{array}\right).
\end{equation}
As long as the number of constraints $M$ is sublinear with respect to $N$ we have that the microcanonical and canonical ensemble studied in subsection $\ref{muc1}$ are equivalent in the thermodynamic limit and $\Sigma\simeq S/N$.
\subsubsection{Multiplex ensemble with given degree sequence in each layer}
 We can fix the the  degree $k_i^{\alpha}$ of every node $i$ in each layer $\alpha$.
 In this case we have $K=M\times N$ constraints  in the system indicated with a labels $\alpha=1,2,\ldots, M$ and $i=1,2\ldots, N$. These constraints are given by 
 \begin{equation}
 F_{i,\alpha}(\vec{G})=k_i^{\alpha}.
 \end{equation}
with  $F_{i,\alpha}(\vec{G})$ given by
 \begin{equation}
 F_{i,\alpha}(\vec{G})=\sum_{j=1,j\neq i}^Na_{ij}^{\alpha}.
 \end{equation}
For this ensemble we can use the results of  \cite{BC,AB2010} getting 
\begin{equation}
N\Sigma=S-N\Omega
\end{equation}
with $S$ given by Eq. $(\ref{Suc2})$ and $N\Omega$ for sparse networks is  given by 
\begin{equation}
N\Omega=-\sum_{\alpha=1}^M\sum_{i=1}^N\ln \pi_{k_i^{\alpha}}(k_i^{\alpha})
\end{equation}
where $\pi_{y}(x)$ is the Poisson distribution with average $y$   $\pi_{y}(x) =1/x! y^x \exp[-y]$. In this case, if the number of layers $M$ is finite, then in the large network limit $N\gg 1$, $\Omega$ is finite, and we have $\Sigma=S/N-\Omega$. Therefore the Gibbs entropy $\Sigma$ is lower than $S/N$  and the microcanonical ensemble is not equivalent in the thermodynamic limit $N\gg 1$ to the conjugated canonical ensemble.
In the case in which $k_i^{\alpha}<\sqrt{\avg{k^{\alpha}}N}$ we can use for $S$ the expression in Eq. $(\ref{Suc2b})$. Therefore the Gibbs entropy $\Sigma $ can be approximated by
\begin{equation}
N\Sigma=\sum_{\alpha=1}^M\log \left[\frac{(\avg{k^{\alpha}}N)!!}{\prod_{i=1}^N k_i^{\alpha}!}e^{-\frac{1}{4}\left(\frac{\Avg{(k^{\alpha})^2}}{\Avg{k^{\alpha}}}\right)^2}\right].
\end{equation}
This last expression is a generalization of the Bender formula \cite{Bender,AB2009} for the entropy of networks with given degree sequence. 
\subsubsection{Multiplex ensemble with given number of links in each layer between nodes of different communities }
 We can fix the total number of links  between nodes of   different communities in each layer $\alpha$.
 We assign to each node $i$ a discrete variable $q_i=1,2,\ldots, Q$ indicating the community of the node.
  We  consider a  microcanonical uncorrelated multiplex ensemble in which we  fix  the total number of links $e_{q,q'}^{\alpha}$ between nodes in community $q$ and nodes in community $q'$ in layer $\alpha$.
 In this case we have $K=M\times Q(Q+1)/2$ constraints  in the system indicated with a labels $\alpha=1,2,\ldots, M$ and $q,q'=1,2\ldots, Q$. These constraints are given by 
 \begin{eqnarray}
 F_{q,q'\alpha}(\vec{G})&=&e_{q,q'}^{\alpha}
 \end{eqnarray} 
where explicit expression for $F_{q,q',\alpha}(\vec{G})$ is given by
 \begin{eqnarray}
 F_{q,q',\alpha}(\vec{G})&=&\sum_{i,j}a_{ij}^{\alpha}\delta_{q,q_i}\delta_{q',q_j}, \ \mbox{for} \  q\neq q' \nonumber \\
 F_{q,q,\alpha}(\vec{G})&=&\sum_{i<j}a_{ij}^{\alpha}\delta_{q,q_i}\delta_{q,q_j}.
 \end{eqnarray}
 
The microcanonical partition function $Z_M$ is equal to the number of multiplexes in this ensemble, which is given by the product over the layers $\alpha=1,2\ldots, M$ of the number of networks $G^{\alpha}$ satisfying the constraints $F_{q,q',\alpha}(\vec{G})=e^{\alpha}_{q,q'}$.  The number of networks $G^{\alpha}$ with $e^{\alpha}_{q,q'}$ links is given by the number of ways of choosing $e_{q,q'}^{\alpha}$ links out of the total number of possible links between nodes in community $q$ and community $q'$, we have, therefore, 
\begin{equation}
Z_M=\prod_{\alpha=1}^M \left[ \prod_{q< q'}\frac{(n_q n_{q'})!}{e_{q,q'}^{\alpha}!}\prod_q \frac{(n_q(n_q-1)/2)!}{e_{q,q}^{\alpha}!}\right],
\end{equation} 
where $n_q$ indicates the number of nodes in community $q$.
 Finally the Gibbs entropy $\Sigma$ for this ensemble is given by Eq. $(\ref{ZM})$ and therefore we obtain 
 \begin{equation}
 N\Sigma=\sum_{\alpha=1}^M \log \left[ \prod_{q< q'}\frac{(n_q n_{q'})!}{e_{q,q'}^{\alpha}!}\prod_q \frac{(n_q(n_q-1)/2)!}{e_{q,q}^{\alpha}!}\right].
 \end{equation}
 In this case the Gibbs entropy $\Sigma= S/N$ in the limit $N\gg 1$ only if the number of constraints $P$ is sublinear with respect to $N$.
 \subsubsection{Multiplex ensemble with given  degree sequence in each layer and given number of links in between nodes in different communities in each layer}
 We assign to each node $i$ a label $q_i=1,2\ldots, Q$ indicating the community to which node $i$ belongs.
 We can consider microcanonical  uncorrelated multiplex ensemble in which we  fix the degree $k_i^{\alpha}$ of every node $i$ in every layer $\alpha$ together with the total  number of links $e_{q,q'}^{\alpha}$ between nodes in community $q$ and nodes in community $q'$ in layer $\alpha$.
 In this case we have $M\times N$ constraints  in the system indicated with a labels $\alpha=1,2,\ldots, M$ and $i=1,2\ldots, N$ and other $M\frac{Q(Q+1)}{2}$ constraints  indicated with labels $\alpha=1,2,\ldots, M$ and $q,q'=1,2\ldots,Q$. These constraints are given by 
 \begin{eqnarray}
F_{i,\alpha}(\vec{G})&=&k_i^{\alpha} \nonumber \\
 F_{q,q',\alpha}(\vec{G})&=&e_{q,q'}^{\alpha}
 \end{eqnarray}
 where the explicit expression for $F_{i,\alpha}(\vec{G})$ and for $F_{q,q',\alpha}(\vec{G})$ is given by
 \begin{eqnarray}
 F_{i,\alpha}(\vec{G})&=&\sum_{j=1,j\neq i}^Na_{ij}^{\alpha}\nonumber \\
 F_{q,q',\alpha}(G^{\alpha})&=&\sum_{i,j}a_{ij}^{\alpha}\delta_{q,q_i}\delta_{q',q_j}, \ \mbox{for} \  q\neq q' \nonumber \\
 F_{q,q,\alpha}(\vec{G})&=&\sum_{i<j}a_{ij}^{\alpha}\delta_{q,q_i}\delta_{q,q_j},
 \end{eqnarray}
 The Gibbs entropy for this ensemble satisfies 
 \begin{equation}
 N\Sigma=S-\sum_{\alpha}N\Omega^{\alpha},
 \end{equation}
 where $S$ is given by Eq. $(\ref{Suc4})$ and using the results of  \cite{AB2010} the entropy of large variations $\Omega^{\alpha}$ for sparse networks is given by 
 \begin{equation}
 N\Omega^{\alpha}=-\sum_{i=1}^N \log \left[ \pi_{k_i^{\alpha}}(k_i^{\alpha}) \right]-\sum_{q\leq q'} \log \left[ \pi_{e_{q,q'}^{\alpha}}(e_{q,q'}^{\alpha}) \right]
 \end{equation}
 where $\pi_{y}(x)$ is the Poisson distribution with average $y$  given by $\pi_{y}(x) =\frac{1}{x!} y^x \exp[-y]$.
   In this case, if the number of layers $M$ is finite, then in the large network limit $N\gg 1$, $\Omega$ is finite, and we have $\Sigma=S/N-\Omega$. Therefore the Gibbs entropy $\Sigma$ is lower than $S/N$  and the microcanonical ensemble is not equivalent in the thermodynamic limit to the conjugated canonical ensemble.
  \subsubsection{Multiplex with given degree-degree correlations in each layer $\alpha$}
We can construct a microcanonical uncorrelated multiplex ensemble with given degree-degree correlations in each layer $\alpha$ by fixing the degree $k_i^{\alpha}$ of each node $i$  in layer $\alpha$ and the total number of links $e_{k,k'}^{\alpha}$ between nodes of degree $k$ and degree $k'$ in layer $\alpha$.
This case is a small modification of the previous case in which for every different layer we identify a community of nodes at a given layer $\alpha$ as the set of nodes with given degree, i.e. $q_i=k_i^{\alpha}$.
The Gibbs entropy $\Sigma$ satisfies
\begin{equation}
 N\Sigma=\sum_{\alpha=1}^MS^{\alpha}-\sum_{\alpha=1}^MN\Omega^{\alpha}.
 \end{equation}
Using the results of  \cite{AB2010} the entropy of large variations $\Omega^{\alpha}$ for sparse networks is given by 
 \begin{equation}
 N\Omega^{\alpha}=-\sum_{i=1}^N \log \left[ \pi_{k_i^{\alpha}}(k_i^{\alpha}) \right]-\sum_{k\leq k'} \log \left[ \pi_{e_{k,k'}^{\alpha}}(e_{q,q'}^{\alpha}) \right].
 \end{equation}
 Moreover the Shannon entropy $S^{\alpha}$ for each layer $\alpha$ is given by 
  \begin{equation}
  S^{\alpha}=-\sum_{i<j} p_{ij}^{\alpha} \log p_{ij}^{\alpha}-\sum_{i<j} (1-p_{ij}^{\alpha}) \log (1-p_{ij}^{\alpha})
  \end{equation}
  with 
  \begin{equation}
  p_{ij}^{\alpha}=\frac{e^{-\lambda_{i,\alpha}-\lambda_{j,\alpha}-\lambda_{k,k',\alpha}}}{1+e^{-\lambda_{i,\alpha}-\lambda_{j,\alpha}-\lambda_{k,k',\alpha}}}
  \end{equation}
  and the Lagrangian multipliers $\lambda_{i,\alpha}$ and $\lambda_{k,k',\alpha}$ fixed by the conditions
  \begin{eqnarray}
  \sum_{j=1,j\neq i}^Np_{ij}^{\alpha}&=&k_i^{\alpha}\nonumber \\
  \sum_{i,j} p_{ij}^{\alpha}\delta_{k_i^{\alpha},k}\delta_{k_j^{\alpha},k'}&=&e_{k,k'}^{\alpha}\ \mbox{for} \ {k\neq k'}\nonumber \\
  \sum_{i<j} p_{ij}^{\alpha}\delta_{k_i^{\alpha},k}\delta_{k_j^{\alpha},k}&=&e_{k,k}^{\alpha}.
  \end{eqnarray}
  \subsection{Correlated microcanonical multiplex ensembles}
  In a correlated multiplex ensemble we have that the probability of a multiplex $\vec{G}$ is not factorizable into the product of probabilities $P_{\alpha}(G^{\alpha})$ of the networks $G^{\alpha}$ in layer $\alpha$, i.e.
\begin{equation}
P_M(\vec{G})\neq\prod_{\alpha=1}^M P_M^{\alpha}(G^{\alpha}).
\label{CCC2}
\end{equation}
The simplest example of correlated multiplex ensemble is the ensemble in which we fix the total number of multilinks $\vec{m}$ in the multiplex.
Starting from this model different more refined multiplex ensemble can be determined, fixing for example the multidegree sequence or the total number of multilinks $\vec{m}$ in between nodes of different communities etc..
In subsection $\ref{mec}$ we will discuss in detail some relevant examples of correlated multiplex ensembles.
\subsection{Examples of correlated microcanonical ensembles}
\label{mec}
\subsubsection{Multiplex ensemble with given total number of multilinks $\vec{m}$}
 In a correlated multiplex ensemble we can fix the total  number $L^{\vec{m}}$ of multilinks  $\vec{m}$ in the multiplex, i.e.
 \begin{equation}
 F_{\vec{m}}(\vec{G})={L^{\vec{m}}}
 \end{equation}
 for all $\vec{m}=(m_1,m_2,\ldots,m_{\alpha},\ldots,m_M)$ with $m_{\alpha}=0,1$, as long as $\sum_{\vec{m}}L^{\vec{m}}=N(N-1)/2$.
 In this case the functions $F_{\vec{m}}(\vec{G})$ are given by 
 \begin{equation}
 F_{\vec{m}}(\vec{G})=\sum_{i<j}A_{ij}^{\vec{m}},
 \end{equation}
  where the multiadjacency matrices of elements $A_{ij}^{\vec{m}}$ are defined as 
 in Eq. (\ref{MA}).
Since any pair of nodes is linked by one multilink $\vec{m}$, we have the total number of multiplexes $Z_M$ in this ensemble is given by the 
multinomial 
\begin{equation}
Z_M=\frac{\left(\begin{array}{c} N\\2 \end{array}\right)!}{\prod_{\vec{m}}L^{\vec{m}}!}.
\end{equation}
Using this result, we can easily derive the Gibbs entropy $N\Sigma=\log(Z_M)$, i.e.
\begin{equation}
N\Sigma=\log \left[\frac{\left(\begin{array}{c} N\\2 \end{array}\right)!}{\prod_{\vec{m}}L^{\vec{m}}!}\right].
\end{equation}

As long as the number of constraints $K=2^M$ is sublinear with respect to $N$ we have that the microcanonical and the conjugated canonical ensemble  are equivalent in the thermodynamic limit  $N\gg 1$ and $\Sigma\simeq S/N$.

\subsubsection{Multiplex ensemble with given multidegree sequence }

In a correlated multiplex ensemble  we can fix the multidegree $k_i^{\vec{m}}$ of node $i$,
 \begin{equation}
 F_{i,\vec{m}}(\vec{G})=k_i^{\vec{m}}
 \end{equation}
 for all $\vec{m}$ with $m_{\alpha}=0,1$ and all $i=1,2,\ldots N$ as long as $\sum_{\vec{m}k_i^{\vec{m}}}=N-1$ and the constraints are graphical. 
 In this case  we have that $F_{i,\vec{m}}(\vec{G})$ is given by 
 \begin{equation}
 F_{i,\vec{m}}(\vec{G})=\sum_{j=1,j\neq i}^NA_{ij}^{\vec{m}},
 \end{equation}
where
 the multiadjacency matrices of elements $A_{ij}^{\vec{m}}=0,1$ are given by Eq. (\ref{MA}). 
 The Gibbs entropy $\Sigma $ of this ensemble satisfies Eq. (\ref{rsso}) that we rewrite here for convenience
\begin{equation}
N\Sigma=S-N\Omega
\end{equation}
with $S$ given by Eq. $(\ref{Skim})$. Using a similar derivation as the one reported in   \cite{BC,AB2010} it is possible to prove that for sparse networks $\Omega$ is given by  
\begin{equation}
N\Omega=-\sum_{\vec{m}|\sum_{\alpha=1}^M m_{\alpha}>0} \sum_{i=1}^N\log \pi_{k_i^{\vec{m}}}(k_i^{\vec{m}})
\end{equation}
where $\pi_{y}(x)$ is the Poisson distribution with average $y$  $\pi_{y}(x)=\frac{1}{x!}y^x\exp[-y]$ calculated at $x$.  
 In this case, if the number of layers $M$ is finite, then in the large network limit $N\gg 1$, $\Omega$ is finite, and we have $\Sigma=S/N-\Omega$. Therefore the Gibbs entropy $\Sigma$ is lower than $S/N$  and the microcanonical ensemble is not equivalent in the thermodynamic limit to the conjugated canonical ensemble.

For networks with $k_i^{\vec{m}}<\sqrt{\avg{k^{\vec{m}}}N}$ where $\vec{m}$ satisfy the  inequality $\sum_{\alpha=1}^M m_{\alpha}>0$, using Eq. $(\ref{Skimu})$ we can find a simple expression for the Gibbs entropy extending Bender result  \cite{Bender, AB2009} to correlated multiplex, i.e.  
\begin{equation}
N\Sigma=\log \left(\prod_{\vec{m}}\frac{(2L^{\vec{m}})!!}{\prod_{i=1}^N k_i^{\vec{m}}!}e^{-\frac{1}{4}\left(\frac{\Avg{(k^{\vec{m}})^2}}{\Avg{k^{\vec{m}}}}\right)^2}\right)
\end{equation}
\subsubsection{Multiplex ensemble with given  number of multilinks $\vec{m}$ in between nodes of different communities}
 We assign to each node $i$ a label $q_i=1,2\ldots, Q$ indicating the community to which node $i$ belongs.
  We  consider a  microcanonical correlated multiplex ensemble in which we  fix  the total number of multilinks $\vec{m}$, $e_{q,q'}^{\vec{m}}$ between nodes in community $q$ and nodes in community $q'$ with the condition that the constraint is graphical.
 In this case we have  $2^M\times \frac{Q(Q+1)}{2}$ constraints  indicated with labels $\vec{m}=(m_1,m_2,\ldots, m_{\alpha},\ldots, m_M)$ with $m_{\alpha}=0,1$ and $q,q'=1,2\ldots,Q$. The constraints
  are given by 
 \begin{eqnarray}
 F_{q,q',\vec{m}}(\vec{G})&=&\sum_{i,j}A_{ij}^{\vec{m}}\delta_{q,q_i}\delta_{q',q_j}=e_{q,q'}^{\vec{m}} \ \mbox{for} \  q\neq q' \nonumber \\
 F_{q,q,\vec{m}}(\vec{G})&=&\sum_{i<j}A_{ij}^{\vec{m}}\delta_{q,q_i}\delta_{q,q_j}=e_{q,q}^{\vec{m}}.
 \end{eqnarray}
For every pair of nodes, one in community $q$ and one in community $q'$ we will have one multilink $\vec{m}$, therefore the total number of multiplex in this ensemble is given by $Z_M$ that has the explicit expression
\begin{equation}
Z_M=\left[ \prod_{q< q'}\frac{(n_q n_{q'})!}{\prod_{\vec{m}}e_{q,q'}^{\vec{m}}!}\prod_q \frac{(n_q(n_q-1)/2)!}{\prod_{\vec{m}}e_{q,q}^{\vec{m}}!}\right],
\end{equation}
where $n_q$ is the number of nodes in community $q$.
Finally the Gibbs entropy $\Sigma$ of this ensemble, with  $N\Sigma=\log Z_M$ satisfies
\begin{equation}
N\Sigma=\log \left[ \prod_{q< q'}\frac{(n_q n_{q'})!}{\prod_{\vec{m}}e_{q,q'}^{\vec{m}}!}\prod_q \frac{(n_q(n_q-1)/2)!}{\prod_{\vec{m}}e_{q,q}^{\vec{m}}!}\right].
\end{equation}
As long as the number of constraints $P$ is sublinear with respect to $N$ we have that the microcanonical and canonical ensemble are equivalent in the thermodynamic limit and $\Sigma\simeq S/N$.
\subsubsection{Multiplex ensemble with given  multidegree sequence and given number of multilinks in between nodes of different communities}
 We assign to each node $i$ a label $q_i=1,2\ldots, Q$ indicating the community to which node $i$ belongs.
 We can consider a microcanonical correlated multiplex ensemble in which we  fix the multidegree $k_i^{\vec{m}}$ of every node $i$  together with the total  number of multilinks $e_{q,q'}^{\vec{m}}$ between nodes in community $q$ and nodes in community $q'$ with the condition that the constraints are graphical.
 In this case we have $2^M\times N$ constraints   indicated with a labels $\vec{m}=(m_1,m_2,\ldots, m_M)$ with $m_{\alpha}=0,1$ and $i=1,2\ldots, N$ and other $2^M\times \frac{Q(Q+1)}{2}$ constraints  indicated with labels $\vec{m}$ and $q,q'=1,2\ldots,Q$. The constraints 
  are given by 
 \begin{eqnarray}
 F_{i,\vec{m}}(\vec{G})&=&\sum_{j=1,j\neq i}^NA_{ij}^{\vec{m}}=k_i^{\vec{m}}\nonumber \\
 F_{q,q',\vec{m}}(\vec{G})&=&\sum_{i,j}A_{ij}^{\vec{m}}\delta_{q,q_i}\delta_{q',q_j}=e_{q,q'}^{\vec{m}} \ \mbox{for} \  q\neq q' \nonumber \\
 F_{q,q,\vec{m}}(\vec{G})&=&\sum_{i<j}A_{ij}^{\vec{m}}\delta_{q,q_i}\delta_{q,q_j}=e_{q,q}^{\vec{m}}.
 \end{eqnarray}
 The Gibbs entropy $\Sigma $ of this ensemble satisfies
 \begin{equation}
 N\Sigma=S-N\Omega
 \end{equation}
 where $S$ is given by Eq. $(\ref{Sck})$ and by following arguments similar to the ones in  \cite{AB2010} it can be proved that for sparse networks $\Omega$ satisfies the following relation
\begin{eqnarray}
N\Omega&=&-\sum_{i=1}^M \sum_{\vec{m}|\sum_{\alpha=1}^Mm_{\alpha}>0}\log \left[ \pi_{k_i^{\vec{m}}}(k_i^{\vec{m}}) \right]\nonumber \\&&-\sum_{q\leq q'} \sum_{\vec{m}|\sum_{\alpha=1}^M m_{\alpha}>0}\log \left[ \pi_{e_{q,q'}^{\vec{m}}}(e_{q,q'}^{\vec{m}}) \right],
\end{eqnarray}
where $\pi_{y}(x)$ is the Poisson distribution with average $y$  $\pi_{y}(x)=\frac{1}{x!}y^x\exp[-y]$ calculated at $x$.  
 In this case, if the number of constraints  $P\propto N$, then in the large network limit $N\gg 1$, $\Omega$ is finite, and we have $\Sigma=S/N-\Omega$. Therefore the Gibbs entropy $\Sigma$ is lower than $S/N$  and the microcanonical ensemble is not equivalent in the thermodynamic limit to the conjugated canonical ensemble.
\section{Conclusions}
In conclusion, we have presented a statistical mechanics approach  for microcanonical and canonical multiplex ensembles.
We have defined  both uncorrelated and correlated multiplex ensembles. Uncorrelated multiplex ensembles are characterized by a probability of the multiplex that factorize into the probability of the networks $G^{\alpha}$ for every layer $\alpha$ of the multiplex.
Therefore for uncorrelated multiplex ensemble the probability a link in one network is independent on the presence of other links in   the other layers. We have considered uncorrelated networks in which we fix the expected number of links in each layer, the expected degree sequence in each layer, the expected number of links in between different communities in each layer, or  the expected degree sequence and the expected total number of links between communities in each layer.
These ensembles, when describing multiplexes  formed by sparse networks, have negligible  global and local  overlap, therefore they cannot model situations in which the overlap of links in different layers is significant.
In order to describe the situation in which the overlap is significant we introduced canonical correlated multiplex ensembles in which we fix the expected  number of multilinks  $\vec{m}$ given by $L^{\vec{m}}$, or the expected multidegree $k_i^{\vec{m}}$ sequence, or the expected number of multilinks $\vec{m}$ between nodes in different communities, or even expected multidegree sequence and expected number of multilinks between nodes of different communities.
Finally we characterize both microcanonical uncorrelated and correlated networks showing that the microcanonical ensembles and canonical ensembles are not equivalent as long as the number of constraints is extensive.
This paper open a new scenario  for studying multiplex ensembles and characterize null models of multiplex including a significant global or local overlap of the links in the different layers. 
In future works we plan to extend this statistical mechanics of multiplex ensembles to more complex situations such as to directed and weighted networks, and to apply the entropy of multiplex  for extracting in formation from multiplex datasets.
Moreover, recently new entropy measures  for quantifying complexity of complex networks have been proposed using tools of quantum in formation theory \cite{Severini,Silvano}. In future works we plan to generalize also these measures to multiplexes and use these new measure to uncover hidden statistical features of multiplex datasets.


\end{document}